\documentclass[letterpaper]{article}

\usepackage[T1]{fontenc}

\usepackage{geometry}
\usepackage{setspace}

\usepackage[
  backend=biber,
  style=chem-acs,   % chem-acs
  sorting=none,
  doi=true,
  url=false,
  isbn=false,
  eprint=false,
  articletitle=true,
  maxbibnames=999
]{biblatex}
\let\cite\supercite

\renewbibmacro{in:}{}

\AtEveryBibitem{%
  \clearfield{issn}%
  \clearfield{url}%
  \clearfield{urldate}%
}

\DeclareSourcemap{
  \maps[datatype=bibtex]{
    \map{
      \step[fieldsource=journaltitle,
            match=\regexp{\s+\d{4}\s+\d+:\d+\s*$},
            replace={}]
    }
    \map{
      \step[fieldsource=journal,
            match=\regexp{\s+\d{4}\s+\d+:\d+\s*$},
            replace={}]
    }
  }
}
\renewbibmacro*{journal+issuetitle}{%
  \usebibmacro{journal}%
  \setunit{\addspace}%
  \usebibmacro{issue+date}%
  \setunit{\addcomma\space}%
  \usebibmacro{volume+number+eid}%
  \newunit}

\AtEveryBibitem{\clearfield{number}}

\usepackage{graphicx}
\usepackage{float}
\newfloat{scheme}{htbp}{los}
\floatname{scheme}{Scheme}
\floatname{chart}{Chart}
\newfloat{graph}{htbp}{loh}

\usepackage{chemformula} % Formulas using \ch{}
\usepackage[version = 4]{mhchem} % Formulas using \ce{}

\PassOptionsToPackage{hidelinks}{hyperref}
\usepackage{hyperref}
\usepackage[nameinlink]{cleveref}
\crefname{figure}{Fig.}{Figs.}   % lowercase \cref
\Crefname{figure}{Fig.}{Figs.}   % capitalized \Cref

\usepackage{float}
\usepackage{comment}

\DeclareUnicodeCharacter{2009}{\,}

\usepackage{authblk}
\author[1,*]{Riccardo Conte}
\author[1]{Lucienne van der Geest}
\author[2]{Minu Sheeja}
\author[2]{Przemyslaw Gawel}
\author[2]{Cina Foroutan-Nejad}
\author[1,*]{Herre S. J. van der Zant}
\affil[1]{Kavli Institute of Nanoscience, Delft University of Technology, Lorentzweg 1, 2628 CJ Delft, The Netherlands.}
\affil[2]{Institute of Organic Chemistry, Polish Academy of Sciences, Kasprzaka 44/52, 01-224, Warsaw, Poland}

\title{Memristive Switches in Rigid Conjugated Single-Molecule Junctions}
\date{*Email: r.conte@tudelft.nl; H.S.J.vanderZant@tudelft.nl}

\begin{document}

\maketitle

\begin{abstract}
Voltage-driven memristive switching has been reported in molecular junctions, yet its microscopic origin often remains elusive. Here, we study three rigid OPE-like derivatives that lack an obvious internal switching pathway using mechanically controlled break junctions (MCBJs) and observe non-volatile, bistable hysteretic IV characteristics at cryogenic temperature. We introduce a quantitative analysis workflow that classifies memristive IVs, clusters the two conductance states, and extracts switching features and stability metrics from repeated measurements at fixed displacement. While all molecules exhibit memristive behavior, stability and hysteresis reproducibility depend strongly on anchoring and connectivity: the linear biphenyl backbone with thiolate (SAc) anchoring shows the most reproducible, predominantly field-driven hysteresis, whereas the meta-phenyl variant with thioether (SMe) anchoring is dominated by stochastic, current-driven events. The resulting conductance statistics point to an extrinsic, mechanically mediated origin involving contact rearrangements, multi-molecule transport, blinking (open--closed) contacts, injection-point shifts, and $\pi$--$\pi$-stacking dimerization.
\end{abstract}

\section*{Keywords}
molecular electronics; memristor; single-molecule junction; mechanically controlled break junction (MCBJ); electric-field-driven switching; extrinsic switching.

%%%%%%%%%%%%%%%%%%%%%%%%%%%%%%%%%%%%%%%%%%%%%%%%%%%%%%%%%%%%%%%%%%%%%
%% Start the main part of the manuscript here.
%%%%%%%%%%%%%%%%%%%%%%%%%%%%%%%%%%%%%%%%%%%%%%%%%%%%%%%%%%%%%%%%%%%%%
\section{Introduction}
\begin{figure}[ht]
\centering
\includegraphics[width=\linewidth]{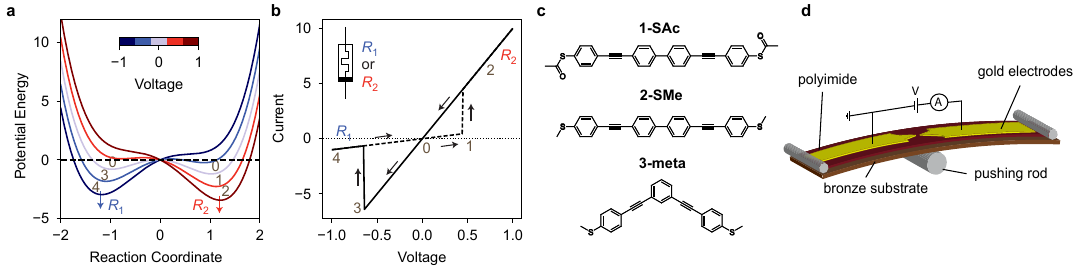}
\caption{(a) Double-well energy landscape of a bistable memristive element with resistance states $R_1$ and $R_2$. The applied bias tilts the landscape, favoring $R_1$ at negative bias (blue) and $R_2$ at positive bias (red). (b) Corresponding current–voltage (IV) response showing a pinched hysteresis loop characteristic of non-volatile bistable switching. (c) Chemical structures of the three rod-like conjugated molecules studied. (d) Schematic of the mechanically controlled break-junction (MCBJ) technique.}
\label{intro}
\end{figure}

Memristive elements, devices whose resistance depends on both the applied bias and the history of electrical stimulation, are emerging as promising building blocks for non-volatile memory, in-memory processing, and neuromorphic hardware \cite{Chua1971Memristor-TheElement, Ielmini2018In-memoryDevices, Kumar2022DynamicalComputing, Xiao2023AProspects}. Their electrical signature is a pinched hysteresis loop in the current–voltage (IV) response. Conceptually, memristive behavior can be viewed as switching between metastable configurations with distinct resistances, whose relative stability is tuned by the applied bias. The bias effectively reshapes the underlying potential landscape, favoring one state over the rest and enabling reproducible switching. Figures~\ref{intro}a--b illustrate the special case of a non-volatile bistable memristor.

Aiming for the ultimate limit of miniaturization, molecular electronics \cite{VonHippel1956MolecularEngineering, Cuevas2010MolecularExperiment, Gorenskaia2024MethodsBehaviour} offers an approach to study and implement such functionality using single molecules as circuit elements \cite{Chen2025ResponsiveDiversification, Bandyopadhyay2006WritingMicroscope, Bandyopadhyay2010MassivelyLayer, Han2020Electric-field-drivenJunctions, Goswami2020AnElectronics, Goswami2021DecisionMemristor, Goswami2017RobustAzoaromatics, Linnenberg2018AddressingStates, Yin2017AAntiaromaticity}. Voltage-driven conductance switching and hysteresis have a long and debated history in molecular junctions. Since the early 2000s, large-area metal/molecular-assembly/metal devices have reported multi-state stability and hysteresis under bias, yet in some cases largely independent of the molecular identity \cite{Stewart2003Molecule-IndependentDevices, Blum2005MolecularlySwitching}. Subsequent work connected many of these ``molecule-independent'' signatures to extrinsic mechanisms associated with electrodes and interfaces, such as atomic rearrangements and filamentary conduction pathways \cite{Lortscher2006ReversibleJunction, Jakobsson2007OnDevices, He2005Metal-freeDevice}. The presence of this appealing functionality and ambiguous microscopic origin triggered extensive efforts to disentangle intrinsic molecular switching from contact-mediated effects using improved fabrication strategies and single-molecule junction techniques \cite{He2005Metal-freeDevice, Troisi2006Molecularmolecular, Prime2009OverviewDevices, JanVanDerMolen2010ChargeSwitches}. However, even single-molecule junctions can exhibit switching driven by the molecule–electrode interface, which is itself a dynamic and bias-sensitive nanoscale system \cite{Leary2015IncorporatingGroup, Gorenskaia2024MethodsBehaviour}. 
Establishing whether a measured switch is intrinsic to the molecular nature or extrinsic to the interface, therefore, remains central both for fundamental interpretation and for assessing device prospects. 

In this contribution, we aim to answer a key question: \textit{
To what extent can non-volatile memristive switching arise in rigid single-molecule junctions that lack an obvious internal switching pathway, and what role do contact-mediated effects play?} We measure conjugated OPE-like derivatives at room-temperature, for their conductance characterization, and at cryogenic temperature for higher stability IV spectroscopy. We observe robust, non-volatile memristive switching that we connect to metal-molecule or molecule-molecule interactions. To systematically probe how anchoring chemistry and molecular connectivity modulate the effect, we study three molecules (Fig.~\ref{intro}c), comparing thioacetate (SAc) with thioether (SMe) anchoring groups and different molecular connectivities. Conductance through these molecules were previously measured as references for quantum interference studies \cite{Soni2020UnderstandingJunctions, Gantenbein2019ExploringDerivatives, Wang2013Conformation-controlledDerivatives, Xie2024RegularlyEffects, Jiang2019TurningConductance}. The absence of an extended, \textit{through-bond} $\pi$-conjugation pathway along the central backbone is a unifying structural motif across the studied molecules. In \textbf{1-SAc} and \textbf{2-SMe}, the core is a biphenyl unit that adopts a pronounced inter-ring twist; this torsion strongly suppresses $\pi$-overlap between the two phenyl rings and therefore limits electronic coupling of the two termini \cite{Venkataraman2006DependenceConformation}. In \textbf{3–meta}, the thioanisole–ethynyl substituents are attached to the central phenyl ring in the 1,3-positions (meta substitution), imposing \textit{cross-conjugated} connectivity rather than a linear conjugation path, again preventing efficient through-bond communication between the two sites \cite{Lambert2015BasicElectronics, RoaldHoffmann1968BenzynesBonds}.

\section{Results}
\noindent Measurements are performed in home-built mechanically controlled break-junction (MCBJ) setups. A lithographically defined Au nanowire on a flexible substrate is repeatedly broken and reformed in a three-point bending geometry (Fig.~\ref{intro}d), enabling the formation of molecular junctions when one or more molecules bridge the Au electrodes.

The target molecules were prepared via a modular, divergent synthesis (see Sections J and K of the Supporting Information). Halogenated precursors of the central cores were first functionalized with trimethylsilylacetylene by Sonogashira cross-coupling to install protected ethynyl linkers. Subsequent desilylation afforded the corresponding terminal alkynes, which were then coupled to protected thiophenol-based anchoring units to yield the final compounds in consistently good yields.  

\begin{figure}[ht]
\centering
\includegraphics[width=0.5\linewidth]{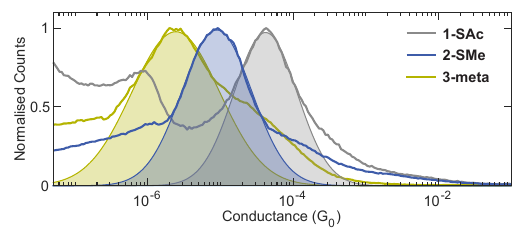}
\caption{Conductance histograms of the three target molecules constructed by overlapping thousands of consecutive room temperature breaking traces (raw data). Gaussian fits, shown as shaded regions, are used to extract the most probable conductance value $G_\mathrm{M}$ (peak position) and the full width at half maximum (FWHM). See Fig.~S1 for the 2D histograms and the measurement details.}
\label{roomT}
\end{figure}

At room temperature, we recorded $5{,}000$ breaking traces at an applied bias of $100\,\mathrm{mV}$ (\textbf{1-SAc} and \textbf{2-SMe}) or $500\,\mathrm{mV}$ (\textbf{3–meta}). Two-dimensional (2D) conductance histograms, formed by superposing the traces, are shown in Fig.~S1. Figure \ref{roomT} displays the raw one-dimensional (1D) histograms, obtained by projecting the data onto the conductance axis. Although the displacement information is lost in this projection, the most probable molecular conductance becomes well defined, and the distribution can be fitted with a Gaussian (shaded regions in \cref{roomT}).  The conductance feature that arises near $8 \cdot10^{-7}\,G_0 $ can be disregarded as it is an instrumentation artifact of room temperature fast-breaking measurements (see SI Sec.~B) . The peak conductances for \textbf{1-SAc}, \textbf{2-SMe}, and \textbf{3–meta} are $4\cdot10^{-5}$, $9\cdot10^{-6}$, and $2\cdot10^{-6}\,\mathrm{G}_{0}$, respectively. These values are consistent with prior reports: for \textbf{1-SAc}: $1.9\cdot10^{-5}\,\mathrm{G}_{0}$ \cite{Wang2013Conformation-controlledDerivatives} and $4.5\cdot10^{-5}\,\mathrm{G}_{0}$ \cite{Soni2020UnderstandingJunctions}; for \textbf{2-SMe}: $2\cdot10^{-5}\,\mathrm{G}_{0}$ \cite{Gantenbein2019ExploringDerivatives}; and for \textbf{3–meta}: $2\cdot10^{-6}\,\mathrm{G}_{0}$ \cite{Xie2024RegularlyEffects}. These room-temperature histogram peaks are interpreted as the average conductance of a fully stretched junction ($G_{\mathrm{M}}$) in which the sulfur anchor binds to Au via thiolate (SAc) or thioether (SMe) coupling. As previously shown for OPE3 systems, the average conductance remains essentially unchanged upon cooling to cryogenic temperatures \cite{Frisenda2018QuantumStudy}. Therefore, the room-temperature conductance values are used as a reference for the fully stretched geometry in the low-temperature analysis. To account for distribution broadening, the full width at half maximum (FWHM) of the Gaussian fits is used to define conductance windows:
$G_{\mathrm{M},\mathrm{SAc}} \in [10^{-5},\,10^{-4}]\,\mathrm{G}_{0}$,
$G_{\mathrm{M},\mathrm{SMe}} \in [2\cdot10^{-6},\,3\cdot10^{-5}]\,\mathrm{G}_{0}$,
and
$G_{\mathrm{M},\mathrm{meta}} \in [4\cdot10^{-7},\,10^{-5}]\,\mathrm{G}_{0}$ (see Fig.~S1 in SI).

Low-temperature measurements were performed in liquid helium ($\sim 6\,\mathrm{K}$) in a high vacuum ($<5\cdot10^{-6}\,\mathrm{mbar}$). For breaking traces, a servomotor actuator was employed instead of a piezo-element (used for room-temperature measurements of \cref{roomT}). An applied bias of $200\,\mathrm{mV}$ was used while breaking. The cryogenic stability of the MCBJ allowed recording IVs up to $1-1.5\,\mathrm{V}$ over multiple sweeps while maintaining a fixed molecular junction, which is not possible at room temperature. 

\begin{figure}[ht]
\centering
\includegraphics[width=\linewidth]{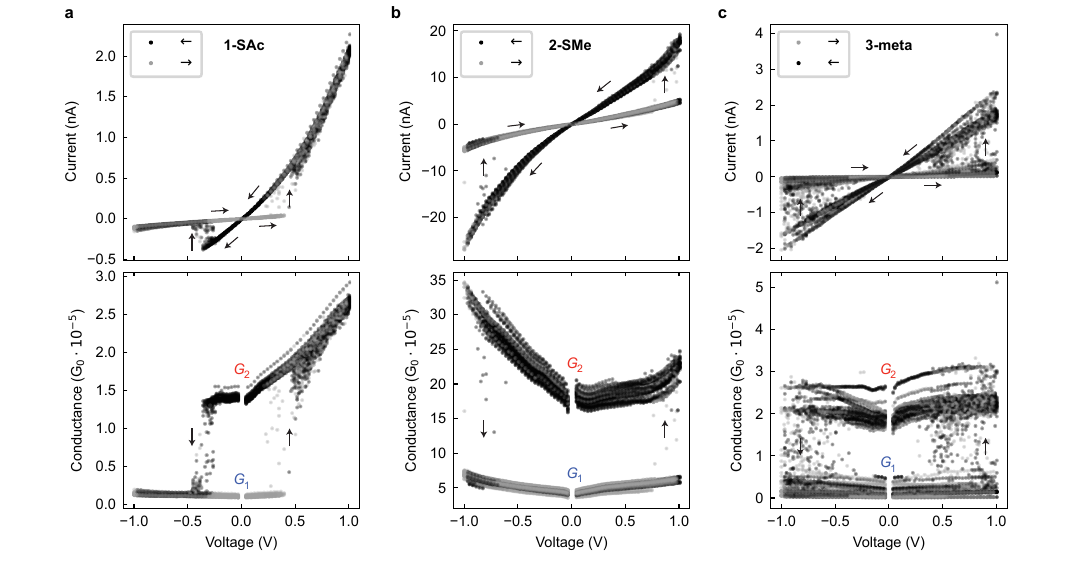}
\caption{Examples of non-volatile memristive IV characteristics recorded at cryogenic temperature. Panels (a--c) show representative IV curves acquired at fixed electrode displacement for each of the three target molecules. $100$ consecutive IVs were recorded to assess switching stability. Colors indicate the sweep direction. The bottom row shows the corresponding conductance–voltage (GV) curves, $G(V)=I/V$.}
\label{memristive_ivs}
\end{figure}

Figure \ref{memristive_ivs} shows non-volatile, bistable memristive IVs observed in fixed-displacement junctions (\emph{hold position}) of the three rod-like molecules at low temperature. Colors denote opposite voltage sweep directions. For each hold position, $100$ IVs were acquired in sequence (each sweep $\sim 1$\,min) to assess switching stability and stochasticity. From each IV, the conductance $G(V)=I/V$ is derived and the conductance–voltage (GV) curve is plotted to more clearly resolve the two states (bottom panels of \cref{memristive_ivs}). Across the three molecules, a broad range of switching behaviors is observed, including different conductance levels, threshold voltages, hysteresis shapes, and nonlinearities (see SI Sec.~I). Despite the absence of an obvious internal structural or chemical switching process, robust memristive behavior emerges in all three systems. 
 
\begin{figure}[ht]
\centering
\includegraphics[width=\linewidth]{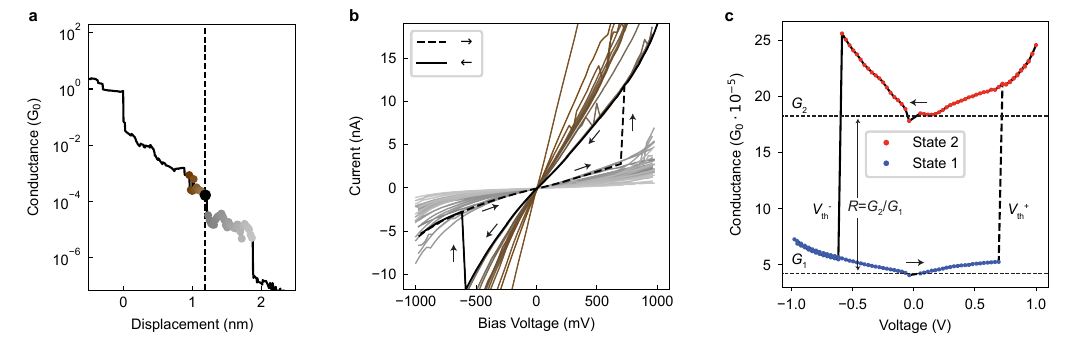}
\caption{Low-temperature protocol for acquiring and analyzing memristive IV characteristics. (a)   Representative molecular breaking trace measured for \textbf{2--SMe}. IV sweeps are acquired at fixed hold positions selected within the conductance window $10^{-3}\!-\!10^{-6}\,\mathrm{G}_{0}$. The vertical dotted line marks the hold position at which a memristive IV is recorded. Brown and gray markers indicate neighboring hold positions at smaller and larger electrode displacement, respectively. (b) Corresponding IV characteristics. The memristive IV is shown in black; solid and dotted lines denote the two voltage sweep directions.   (c)  Clustering of the corresponding conductance--voltage data into two conductance states. As in \cref{intro}a--b, red denotes the high-conductance state and blue the low-conductance state. The features extracted for the statistical analysis are indicated in the figure: zero-bias conductances $G_{1}$ and $G_{2}$, their ratio $R = G_{2}/G_{1}$, and switching thresholds $V_{\mathrm{th}}^{+}$ and $V_{\mathrm{th}}^{-}$.}
\label{features}
\end{figure}

Figure \ref{features} summarizes the masurement and analysis workflow that we developed to identify switching events and extract their features (details in SI, Sec.~C). For each breaking trace (\cref{features}a), IVs from $-1$ to $+1\,\mathrm{V}$ are recorded at multiple hold positions (\cref{features}b). These hold positions are selected between the rupture of the Au bridge and the amplifier noise floor, corresponding to a typical conductance range from $10^{-2}$--$10^{-3}$ down to $10^{-6}\,\mathrm{G}_{0}$. Each IV is classified as \emph{memristive} or \emph{non-memristive} according to the presence of bistable, non-volatile hysteretic switching. In \cref{features}b, the memristive hold position is shown in black, while the neighboring hold positions at a smaller and larger electrode displacement are shown in brown and gray, respectively. The classification procedure is illustrated in \cref{features}c. The conductance--voltage data (GV) are clustered into two states and memristivity is evaluated using criteria that verify (i) clear conductance separation, (ii) the existence of both states at low bias (non-volatility), and (iii) opposite switching polarities (one state set at positive bias and the other at negative).  For memristive IVs, the following features are extracted: zero-bias conductances $G_{1}$ and $G_{2}$, their ratio $R=G_{2}/G_{1}$, and switching thresholds $V_{\mathrm{th}}^{+}$ and $V_{\mathrm{th}}^{-}$. Then, an additional $99$ IVs are acquired for the same hold position to assess switching stability. A hold position is deemed \emph{stable} if $\geq 20\%$ of its $100$ IVs are classified as memristive. We refer to the full set of IVs measured at one fixed displacement as an \emph{IV collection}. Figure \ref{memristive_ivs}b shows the IV collection corresponding to the hold position highlighted in \cref{features}.

Approximately $8{,}000$ breaking traces were recorded for \textbf{1–SAc} and $1{,}500$ for \textbf{2–SMe} and \textbf{3–meta}. The molecular fraction, defined as traces with at least three hold positions, was comparable across molecules ($25$–$44\%$). In total, we examined $\sim 20{,}000$ hold positions for \textbf{1-SAc} and $\sim 5{,}000$ for both \textbf{2-SMe} and \textbf{3–meta} (full statistics in SI, Sec.~D). For \textbf{1-SAc}, $16\%$ of molecular traces contained at least one hold position with a memristive IV, and $10\%$ of them were stable by the above criterion. While \textbf{2-SMe} and \textbf{3–meta} yielded a higher number of memristive traces overall (20.5\% and 28\%, respectively), \textbf{1-SAc} yielded the largest fraction of stable traces (10\% versus 5\% and 7\% for \textbf{2-SMe} and \textbf{3–meta}, respectively).  Taken together, these trends indicate that bent (meta) connectivity is more prone to stochastic switching (see, for example, the IV collection in \cref{memristive_ivs}c), whereas linear connectivity reduces switching. Furthermore, when comparing \textbf{1-SAc} with \textbf{2-SMe}, the number of stable hold positions indicates that stronger SAc anchoring promotes more reproducible hysteresis.
 
\begin{figure}[ht]
\centering
\includegraphics[width=\linewidth]{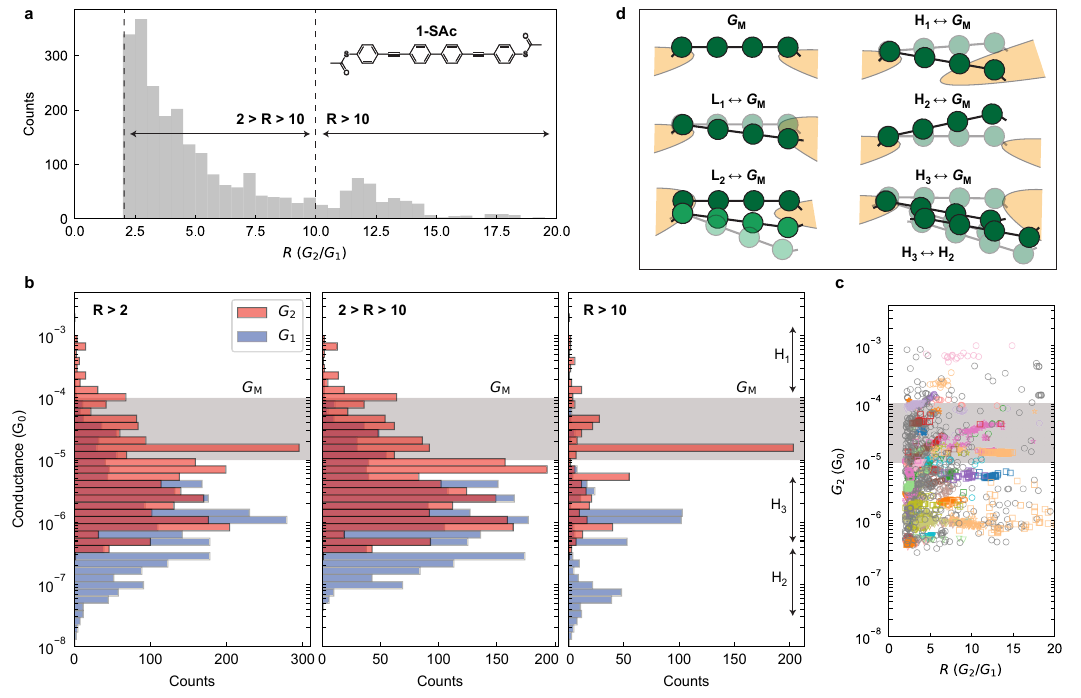}
\caption{Conductance statistics for \textbf{1–SAc}. (a) Distribution of the conductance ratio $R = G_{2}/G_{1}$ across all memristive IVs. A minimum separation $R\geq2$ is enforced in the memristivity classification (see SI Sec.~C). (b) Histograms of the two conductance states ($G_{1}$ and $G_{2}$) grouped by ratio intervals: $R>2$ (all memristive IVs), $2<R<10$ (intermediate conductance separation), and $R>10$ (large separation). (c) Scatter plot of $G_{2}$ versus $R$. Each point corresponds to a memristive IV with marker shape and color identifying the associated IV collection. (d) Schematic of proposed switching mechanisms, grouped into low-ratio (L) and high-ratio (H) families: ($G_{\mathrm{M}}$) fully stretched single-molecule junction; (L$_1\leftrightarrow G_{\mathrm{M}}$) contact rearrangement at Au–anchor; (L$_2\leftrightarrow G_{\mathrm{M}}$) parallel multi-molecule transport; (H$_1\leftrightarrow G_{\mathrm{M}}$) long–short injection switching; (H$_2\leftrightarrow G_{\mathrm{M}}$) blinking (open–closed) contact; (H$_3\leftrightarrow G_{\mathrm{M}}$ or H$_2$) $\pi$–$\pi$ dimer $\leftrightarrow$ single-molecule/open.}
\label{conductance_histo_SAc}
\end{figure}

Figure \ref{conductance_histo_SAc} summarizes the main statistical conductance results for memristive IVs of \textbf{1–SAc} (see Figs.~S6 and S7 for \textbf{2–SMe} and \textbf{3–meta}, respectively). 
Figure \ref{conductance_histo_SAc}a shows the distribution of the zero-bias conductance ratio $R=G_{2}/G_{1}$. The histogram peaks near $3$, decreases with increasing $R$, and exhibits a local maximum around $12$. To further resolve the switching behavior, \Cref{conductance_histo_SAc}b reports histograms of $G_{1}$ and $G_{2}$ for three ratio ranges: $R>2$ (all memristive IVs), $2<R<10$ (low to intermediate ratios), and $R>10$ (highly separated states). The grey band ($G_{\mathrm{M}}$) marks the conductance window of the fully stretched junction inferred from room-temperature measurements (cf. \cref{roomT}), as described earlier. For $2<R<10$, the conductance states are broadly distributed over $2$–$3$ orders of magnitude, with $G_{1}$ centered near $10^{-6}\,\mathrm{G}_{0}$ and $G_{2}$ near $10^{-5}\,\mathrm{G}_{0}$. In contrast, for $R>10$ the distributions sharpen: $G_{1}$ peaks around $5\cdot10^{-8}$ and $10^{-6}\,\mathrm{G}_{0}$, while $G_{2}$ shows a dominant peak at $10^{-5}\,\mathrm{G}_{0}$ and a minor presence near $10^{-6}\,\mathrm{G}_{0}$. Figure \ref{conductance_histo_SAc}c provides a complementary view as a scatter plot of $G_{2}$ versus $R$. Each point corresponds to a memristive IV. Marker shape and color identify its IV collection, and grey points denote collections with fewer than $10$ memristive IVs. This plot highlights the conductance states of stable, high-ratio hold positions that contribute to the dominant peaks in \Cref{conductance_histo_SAc}b.

Note that all memristive IVs are included in \Cref{conductance_histo_SAc}, even when multiple IVs originate from the same collection. Consequently, the visualization effectively weights the switching population by stability. This choice highlights stable hold positions and de-emphasizes sporadic switches. A similar analysis for \textbf{2–SMe} and \textbf{3–meta} yields spikier histograms due to the smaller datasets. Overall, \textbf{2–SMe} resembles \textbf{1–SAc} but with an additional, more pronounced peak near $10^{-4}\,\mathrm{G}_{0}$ (Fig. S6). For \textbf{3–meta}, $G_{1}$ exhibits three peaks around $10^{-7}$, $10^{-6}$, and $10^{-5}\,\mathrm{G}_{0}$, while $G_{2}$ peaks near $10^{-6}$, $10^{-5}$, and $10^{-4}\,\mathrm{G}_{0}$ (Fig. S7).

\section{Discussion}
Based on the low-temperature conductance analysis discussed above, the room-temperature $G_{\mathrm{M}}$ values, and prior literature, we propose several hypotheses for the microscopic origin of the observed switching. Given the absence of an obvious intrinsic switching degree of freedom in these rigid molecules, we attribute the behavior primarily to mechanically driven processes within the junction, involving one or more molecules and the electrodes. The proposed scenarios are summarized schematically in \cref{conductance_histo_SAc}d. The left column of \cref{conductance_histo_SAc}d shows geometries associated with the fully stretched molecular junction $G_{\mathrm{M}}$ and with \emph{low-ratio} events ($R<5$). These events are plausibly driven by rearrangements at the Au–molecule interface (L$_1\leftrightarrow G_{\mathrm{M}}$), such as changes in the number of Au atoms involved in binding or in the local contact geometry \cite{Batista2007ResistiveNanocontacts, Wang2012Voltage-dependentJunction, Grigoriev2006CriticalJunctions, Li2007ChargeFreedom, Gorenskaia2024MethodsBehaviour}. They may also reflect variations in the effective anchoring dipole (or local electrostatic environment) under bias \cite{Olavarria-Contreras2018Electric-fieldCompounds}. An alternative explanation that cannot be excluded is parallel transport through more than one molecule, with small structural rearrangements within a multi-molecule junction (L$_2\leftrightarrow G_{\mathrm{M}}$).

\emph{High-ratio} switching can be grouped into three categories, summarized in the right column of \cref{conductance_histo_SAc}d. The black arrows in \cref{conductance_histo_SAc}b (panel $R>10$) link them to the corresponding conductance features in the histograms; representative IV characteristics for each category are shown in SI Sec.~I. We outline the three categories below:

(H$_1\leftrightarrow G_{\mathrm{M}}$) \emph{Long–short injection switching.} The effective injection point at one electrode toggles to the molecule backbone, giving $G_{1}\!\sim\!G_{\mathrm{M}}$ and a higher $G_{2}$. This behavior is more pronounced for \textbf{2–SMe} and \textbf{3–meta}, where $G_{2}$ shows a marked peak near $10^{-4}\,\mathrm{G}_{0}$ (see SI).  Figure~\ref{features}a-b illustrates an example of this behavior with IVs measured at neighboring hold positions. The high-conductance state is found at smaller electrode displacement (brown), whereas the low-conductance state persists near $10^{-5}\,\mathrm{G}_{0}$ over a wide displacement range (gray) before the conductance falls to the noise floor, consistent with a fully stretched single-molecule junction. Although this mechanism has not previously been identified from switching IVs, earlier studies have shown that electric fields similar to those generated here can stabilize shorter junction geometries \cite{Gao2021ElectricJunctions,Miguel2015TowardDerivatives,Yu2020ChargeOligomers,Tang2020Electric-Field-InducedJunctions}. In those cases, the effect was observed for molecular backbones containing nitrogen heteroatoms. In the present all-carbon molecules, such metastable geometries may become observable because they are stabilized at low temperature. 

(H$_2\leftrightarrow G_{\mathrm{M}}$) \emph{Blinking (open–closed) .} Intermittent contact yields $G_{2}\!\sim\!G_{\mathrm{M}}$ and $G_{1}$ near the noise floor ($\sim10^{-7}\,\mathrm{G}_{0}$), consistent with the low $G_{1}$ peak around $5\cdot10^{-8}\,\mathrm{G}_{0}$ in \cref{conductance_histo_SAc}b. Blinking is well documented in scanning tunneling microscopy (STM) studies \cite{Haiss2004MeasurementWires, Gerhard2017AnSwitch} and has also been proposed for room-temperature MCBJ fast-breaking traces \cite{Ornago2022SwitchingReconfiguration}.

(H$_3\leftrightarrow G_{\mathrm{M}}$ or H$_2$) \emph{$\pi$–$\pi$ stacked dimer $\leftrightarrow$ single-molecule (or open)}. A dimer of two $\pi$-stacked molecules with each connecting an electrode has a conductance between $G_{\mathrm{M}}$ and the open-junction level. Transitions to a single-molecule junction produce a jump toward $G_{\mathrm{M}}$ (or towards H$_2$ when open). The feature near $\sim\!10^{-6}\,\mathrm{G}_{0}$ in \cref{conductance_histo_SAc}b can be compatible with this picture because $\pi$-stacked dimers are typically $1$–$2$ orders of magnitude below $G_{\mathrm{M}}$ \cite{Frisenda2016MechanicallyDimers}, and electric fields can affect the probability of stacking \cite{Tang2020ElectricJunctions}.

The microscopic interpretation of switching events proposed above is based primarily on the zero-bias conductances of the two states ($G_1$ and $G_2$). However, additional information can, in principle, be extracted from the shape of the measured IVs. As a first step in this direction, we analyzed the collections shown in Fig.~4 using a single-level Landauer model (SI Sec.~H). The two states were fitted separately, and only the low-bias segments preceding the switching event were considered. The model contains three free parameters: the zero-bias level position $\varepsilon_0$, the total level broadening $\Gamma$, and the electrostatic asymmetry parameter $\eta$. Although this minimal description is not sufficient to identify the microscopic switching mechanism, the extracted parameters show informative trends. States assigned to stretched or open junction geometries tend to yield $\eta\sim 0$, consistent with comparatively symmetric configurations, whereas states assigned to shorter or locally rearranged junctions show larger $|\eta|$, indicating increased junction asymmetry. In addition, configurations interpreted as closed molecular junctions tend to exhibit smaller level misalignment $\varepsilon_0$ than open configurations. Together, these trends support the view that the observed switching is associated with changes in coupling and electrostatic asymmetry.

The discussion above addresses which conductance states participate in the switching. We now turn to the distinct question of why the transitions between those states occur with a reproducible hysteresis polarity, i.e., what makes the switches memristive. Bias can trigger two broad classes of switching: (i) \emph{current-driven} switching, where exceeding a threshold current induces stochastic telegraph-like fluctuations, and (ii) \emph{field-driven} switching, where the electric field shifts the relative energy of two states \cite{JanVanDerMolen2010ChargeSwitchesb, Liljeroth2007Current-inducedMolecules, Gerhard2017AnSwitch}. Class (i) typically lacks a reproducible hysteresis polarity, whereas class (ii) can yield repeatable IV loops with consistent high-bias states. Field-driven behavior requires that the two states differ in dipole moment and/or polarizability \cite{Qiu2004MechanismsMolecule}, which may arise from changes in the longitudinal dipole \cite{Qiu2004MechanismsMolecule,Alemani2006ElectricSTM,Olavarria-Contreras2018Electric-fieldCompounds}, field-induced torque in a transverse dipole \cite{Gerhard2017AnSwitch, Tanimoto2016DipoleJunctions}, or field-assisted dipole formation/rupture at the molecule-electrode interface \cite{Peng2009ConductanceAlignment, Li2015ElectricJunctions}.

To compare how closely the three molecules approach the ideal field-driven memristive limit, we consider three metrics: (a) the average number of switching events per unidirectional sweep segment (SI Fig.~S5), (b) the threshold-voltage variability within a given IV collection (SI Fig.~S9), and (c) the fraction of stable memristive collections in the full dataset (SI Fig.~S4). In the ideal field-driven case, each sweep direction should contain a single switching event, and the threshold voltage should be highly reproducible across the IVs of the same collection. By these criteria, \textbf{1--SAc} and \textbf{2--SMe} both exhibit predominantly field-driven behavior, with broadly similar characteristics, in contrast to \textbf{3--meta}, which is more stochastic and shows a larger threshold variability. The similarity between \textbf{1--SAc} and \textbf{2--SMe} suggests that the origin of the field-driven switching is linked, at least in part, to the linear molecular geometry and cannot be attributed solely to the anchoring group. The main difference between the two linear molecules is that \textbf{1--SAc} exhibits a higher fraction of stable collections than \textbf{2--SMe}, suggesting that the stronger anchoring group improves the reproducibility of repeated IVs. 

Within the potential-energy-landscape picture of \cref{intro}a, these observations suggest that molecular connectivity primarily governs how the relative energies of the states evolve under bias, whereas the anchoring group mainly affects the number and stability of the accessible local minima. A more quantitative separation of the roles of connectivity, anchoring group, contact geometry, and dipolar response will require DFT calculations over a broad ensemble of geometries and fields, which is beyond the scope of this work. Such an analysis could also clarify to what extent intrinsic molecular effects, such as bias-induced redox processes or subtle conformational changes (e.g., phenyl torsion), contribute alongside the extrinsic mechanisms discussed here.

\section{Conclusion}
\noindent In summary, rigid, rod-like OPE derivatives exhibit bistable, non-volatile memristive switching, which is interpreted as mechanical rearrangements of the junction. Such mechanisms can be exploited for functional devices and must be accounted for when isolating intrinsic molecular processes. Our results raise concerns that some electric-field–induced switches previously ascribed to molecular properties may instead reflect extrinsic mechanics. Mechanical switching occurs in both linear (para) and bent (meta) geometries and with both stronger and weaker anchoring groups (SAc and SMe, respectively). The microscopic picture, derived from statistics of the zero-bias conductances ($G_{1}$, $G_{2}$) and their ratio ($G_{2}/G_{1}$) across all memristive IVs, indicates: (i) rearrangement of Au atoms at the electrodes; (ii) orientation changes in multi-molecule junctions; (iii) shifts of the injection point along the backbone; (iv) intermittent opening/closing of the molecule–electrode contact; and (v) transitions between $\pi$–$\pi$ stacked dimers and single-molecule (or open) junctions. DFT calculations would provide valuable insights to further establish the nature of non-volatile memristive switches in rigid molecules.

\section{Acknowledgements}
M. S. and C.F.N. acknowledge the National Science Centre, Poland (OPUS 2020/39/B/ST4/02022) for funding this work. P. G. acknowledges funding from the National Agency for Academic Exchange (Polish Returns PPN\_PPO\_2020\_1\_00012).
\section{Conflicts of interest}
There are no conflicts to declare.

\section*{Author Contributions}
R.C.: Measurements; Analysis; Data interpretation; Writing---original draft.
L.v.d.G.: Measurements; Formal analysis.
M.S.: Synthesis of molecules.
P.G.: Supervision (synthesis). 
C.F.N.: Supervision; Data interpretation; Writing---review \& editing. 
H.S.J.v.d.Zant: Supervision; Data interpretation; Writing---review \& editing. 
All authors reviewed and approved the final manuscript.

\section*{Supporting information}
The following files are available free of charge.
\begin{itemize}
  \item Supporting\_Information.pdf: Experimental details; room-temperature characterization; low-temperature workflow for identifying memristive IVs; global statistics of memristive IVs; additional conductance histograms; threshold-voltage distributions; representative IV collections for the three molecules; synthetic protocols; and NMR spectra.
\end{itemize}

\section*{Data and Code Availability}
The data and analysis code underlying this study are openly available at https://doi.org/10.5281/zenodo.18593437.

%%%%%%%%%%%%%%%%%%%%%%%%%%%%%%%%%%%%%%%%%%%%%%%%%%%%%%%%%%%%%%%%%%%%%
%% If you are using classical BibTeX rather than biblatex,
%% remove the \printbibliography and uncomment the \bibliograpy one
%%%%%%%%%%%%%%%%%%%%%%%%%%%%%%%%%%%%%%%%%%%%%%%%%%%%%%%%%%%%%%%%%%%%%
\printbibliography

@article{Yin2017AAntiaromaticity,
    title = {{A reversible single-molecule switch based on activated antiaromaticity}},
    year = {2017},
    journal = {Science Advances},
    author = {Yin, Xiaodong and Zang, Yaping and Zhu, Liangliang and Low, Jonathan Z. and Liu, Zhen Fei and Cui, Jing and Neaton, Jeffrey B. and Venkataraman, Latha and Campos, Luis M.},
    number = {10},
    volume = {3},
    publisher = {American Association for the Advancement of Science},
    url = {/doi/pdf/10.1126/sciadv.aao2615?download=true},
    doi = {10.1126/SCIADV.AAO2615},
    issn = {23752548},
    pmid = {29098181}
}

@article{Xiao2023AProspects,
    title = {{A review of memristor: material and structure design, device performance, applications and prospects}},
    year = {2023},
    journal = {Science and Technology of Advanced Materials},
    author = {Xiao, Yongyue and Jiang, Bei and Zhang, Zihao and Ke, Shanwu and Jin, Yaoyao and Wen, Xin and Ye, Cong},
    number = {1},
    pages = {2162323},
    volume = {24},
    publisher = {Taylor and Francis Ltd.},
    url = {https://pmc.ncbi.nlm.nih.gov/articles/PMC9980037/},
    doi = {10.1080/14686996.2022.2162323},
    issn = {18785514},
    pmid = {36872944}
}

@article{Martin2011AJunctions,
    title = {{A versatile low-temperature setup for the electrical characterization of single-molecule junctions}},
    year = {2011},
    journal = {Review of Scientific Instruments},
    author = {Martin, Christian A. and Smit, Roel H.M. and Egmond, Ruud Van and Van Der Zant, Herre S.J. and Van Ruitenbeek, Jan M.},
    number = {5},
    month = {5},
    volume = {82},
    doi = {10.1063/1.3593100},
    issn = {00346748}
}

@article{Linnenberg2018AddressingStates,
    title = {{Addressing Multiple Resistive States of Polyoxovanadates: Conductivity as a Function of Individual Molecular Redox States}},
    year = {2018},
    journal = {Journal of the American Chemical Society},
    author = {Linnenberg, Oliver and Moors, Marco and Notario-Est{\'{e}}vez, Almudena and L{\'{o}}pez, Xavier and De Graaf, Coen and Peter, Sophia and Baeumer, Christoph and Waser, Rainer and Monakhov, Kirill Yu},
    number = {48},
    month = {12},
    pages = {16635--16640},
    volume = {140},
    publisher = {American Chemical Society},
    url = {/doi/pdf/10.1021/jacs.8b08780?ref=article_openPDF},
    doi = {10.1021/JACS.8B08780},
    issn = {15205126},
    pmid = {30418764}
}

@article{Hortholary2003AnMonolayers,
    title = {{An Approach to Long and Unsubstituted Molecular Wires:  Synthesis of Redox-Active, Cationic Phenylethynyl Oligomers Designed for Self-Assembled Monolayers}},
    year = {2003},
    journal = {Journal of Organic Chemistry},
    author = {Hortholary, Cédric and Coudret, Christophe},
    number = {6},
    month = {3},
    pages = {2167--2174},
    volume = {68},
    publisher = { American Chemical Society },
    url = {/doi/pdf/10.1021/jo026735z?ref=article_openPDF},
    doi = {10.1021/JO026735Z},
    issn = {00223263}
}

@article{Gerhard2017AnSwitch,
    title = {{An electrically actuated molecular toggle switch}},
    year = {2017},
    journal = {Nature Communications 2017 8:1},
    author = {Gerhard, Lukas and Edelmann, Kevin and Homberg, Jan and Val{\'{a}}{\v{s}}ek, Michal and Bahoosh, Safa G. and Lukas, Maya and Pauly, Fabian and Mayor, Marcel and Wulfhekel, Wulf},
    number = {1},
    month = {3},
    pages = {14672-},
    volume = {8},
    publisher = {Nature Publishing Group},
    url = {https://www.nature.com/articles/ncomms14672},
    doi = {10.1038/ncomms14672},
    issn = {2041-1723},
    pmid = {28276442}
}

@article{Goswami2020AnElectronics,
    title = {{An organic approach to low energy memory and brain inspired electronics}},
    year = {2020},
    journal = {Applied Physics Reviews},
    author = {Goswami, Sreetosh and Goswami, Sreebrata and Venkatesan, T.},
    number = {2},
    month = {6},
    volume = {7},
    publisher = {American Institute of Physics Inc.},
    url = {/aip/apr/article/7/2/021303/997523/An-organic-approach-to-low-energy-memory-and-brain},
    doi = {10.1063/1.5124155},
    issn = {19319401}
}

@article{Lambert2015BasicElectronics,
    title = {{Basic concepts of quantum interference and electron transport in single-molecule electronics}},
    year = {2015},
    journal = {Chemical Society Reviews},
    author = {Lambert, C. J.},
    number = {4},
    month = {2},
    pages = {875--888},
    volume = {44},
    publisher = {The Royal Society of Chemistry},
    url = {https://pubs.rsc.org/en/content/articlehtml/2015/cs/c4cs00203b https://pubs.rsc.org/en/content/articlelanding/2015/cs/c4cs00203b},
    doi = {10.1039/C4CS00203B},
    issn = {1460-4744}
}

@article{RoaldHoffmann1968BenzynesBonds,
    title = {{Benzynes, dehydroconjugated molecules, and the interaction of orbitals separated by a number of intervening sigma bonds}},
    year = {1968},
    journal = {Journal of the American Chemical Society},
    author = {{Roald Hoffmann} and {Akira Imamura} and {Warren J. Hehre}},
    number = {6},
    month = {3},
    pages = {1499--1509},
    volume = {90},
    publisher = {American Chemical Society},
    url = {/doi/pdf/10.1021/ja01008a018?ref=article_openPDF},
    doi = {10.1021/JA01008A018},
    issn = {15205126}
}

@article{Yu2020ChargeOligomers,
    title = {{Charge Transport in Sequence-Defined Conjugated Oligomers}},
    year = {2020},
    journal = {Journal of the American Chemical Society},
    author = {Yu, Hao and Li, Songsong and Schwieter, Kenneth E. and Liu, Yun and Sun, Boran and Moore, Jeffrey S. and Schroeder, Charles M.},
    number = {10},
    month = {3},
    pages = {4852--4861},
    volume = {142},
    publisher = {American Chemical Society},
    url = {/doi/pdf/10.1021/jacs.0c00043?ref=article_openPDF},
    doi = {10.1021/JACS.0C00043},
    issn = {15205126},
    pmid = {32069403}
}

@article{Li2007ChargeFreedom,
    title = {{Charge Transport in Single Au | Alkanedithiol | Au Junctions:  Coordination Geometries and Conformational Degrees of Freedom}},
    year = {2007},
    journal = {Journal of the American Chemical Society},
    author = {Li, Chen and Pobelov, Ilya and Wandlowski, Thomas and Bagrets, Alexei and Arnold, Andreas and Evers, Ferdinand},
    number = {1},
    month = {1},
    pages = {318--326},
    volume = {130},
    publisher = { American Chemical Society },
    url = {/doi/pdf/10.1021/ja0762386?ref=article_openPDF},
    doi = {10.1021/JA0762386},
    issn = {00027863},
    pmid = {18076172}
}

@article{JanVanDerMolen2010ChargeSwitches,
    title = {{Charge transport through molecular switches}},
    year = {2010},
    journal = {Journal of physics. Condensed matter : an Institute of Physics journal},
    author = {Jan Van Der Molen, Sense and Liljeroth, Peter},
    number = {13},
    volume = {22},
    publisher = {J Phys Condens Matter},
    url = {https://pubmed.ncbi.nlm.nih.gov/21389503/},
    doi = {10.1088/0953-8984/22/13/133001},
    issn = {1361-648X},
    pmid = {21389503}
}

@article{JanVanDerMolen2010ChargeSwitchesb,
    title = {{Charge transport through molecular switches}},
    year = {2010},
    journal = {Journal of Physics: Condensed Matter},
    author = {Jan Van Der Molen, Sense and Liljeroth, Peter},
    number = {13},
    month = {3},
    pages = {133001},
    volume = {22},
    publisher = {IOP Publishing},
    url = {https://iopscience.iop.org/article/10.1088/0953-8984/22/13/133001 https://iopscience.iop.org/article/10.1088/0953-8984/22/13/133001/meta},
    doi = {10.1088/0953-8984/22/13/133001},
    issn = {0953-8984}
}

@article{Perrin2015ChargeTheory,
    title = {{Charge Transport Through Single-Molecule Junctions Experiments And Theory}},
    year = {2015},
    journal = {Delft University of Technology, PhD thesis},
    author = {Perrin, Mickael}
}

@article{Ornago2023ComplexityJunctions,
    title = {{Complexity of Electron Transport in Nanoscale Molecular Junctions}},
    year = {2023},
    journal = {Delft University of Technology, PhD thesis},
    author = {Ornago, L},
    url = {https://doi.org/10.4233/uuid:e6002163-58f0-48e1-bce2-},
    doi = {10.4233/uuid:e6002163-58f0-48e1-bce2-02ac111a8adf}
}

@article{Peng2009ConductanceAlignment,
    title = {{Conductance of Conjugated Molecular Wires: Length Dependence, Anchoring Groups, and Band Alignment}},
    year = {2009},
    journal = {Journal of Physical Chemistry C},
    author = {Peng, Guowen and Strange, Mikkel and Thygesen, Kristian S. and Mavrikakis, Manos},
    number = {49},
    pages = {20967--20973},
    volume = {113},
    publisher = { American Chemical Society},
    url = {/doi/pdf/10.1021/jp9084603?ref=article_openPDF},
    doi = {10.1021/JP9084603},
    issn = {19327447}
}

@article{Wang2013Conformation-controlledDerivatives,
    title = {{Conformation-controlled electron transport in single-molecule junctions containing oligo(phenylene ethynylene) derivatives}},
    year = {2013},
    journal = {Chemistry - An Asian Journal},
    author = {Wang, Le Jia and Yong, Ai and Zhou, Kai Ge and Tan, Lin and Ye, Jian and Wu, Guo Ping and Xu, Zhu Guo and Zhang, Hao Li},
    number = {8},
    month = {8},
    pages = {1901--1909},
    volume = {8},
    publisher = {John Wiley {\&} Sons, Ltd},
    url = {/doi/pdf/10.1002/asia.201300264 https://onlinelibrary.wiley.com/doi/abs/10.1002/asia.201300264 https://aces.onlinelibrary.wiley.com/doi/10.1002/asia.201300264},
    doi = {10.1002/ASIA.201300264},
    issn = {18614728}
}

@article{Grigoriev2006CriticalJunctions,
    title = {{Critical roles of metal-molecule contacts in electron transport through molecular-wire junctions}},
    year = {2006},
    journal = {Physical Review B},
    author = {Grigoriev, A. and Sk{\"{o}}ldberg, J. and Wendin, G. and Crljen, Ž},
    number = {4},
    month = {7},
    pages = {045401},
    volume = {74},
    publisher = {American Physical Society},
    url = {https://journals.aps.org/prb/abstract/10.1103/PhysRevB.74.045401},
    doi = {10.1103/PhysRevB.74.045401},
    issn = {10980121}
}

@article{Liljeroth2007Current-inducedMolecules,
    title = {{Current-induced hydrogen tautomerization and conductance switching of naphthalocyanine molecules}},
    year = {2007},
    journal = {Science},
    author = {Liljeroth, Peter and Repp, Jascha and Meyer, Gerhard},
    number = {5842},
    month = {8},
    pages = {1203--1206},
    volume = {317},
    publisher = {American Association for the Advancement of Science},
    url = {/doi/pdf/10.1126/science.1144366?download=true},
    doi = {10.1126/SCIENCE.1144366;SUBPAGE:STRING:FULL},
    issn = {00368075}
}

@article{Olavarria-Contreras2016CAuGroups,
    title = {{C–Au Covalently Bonded Molecular Junctions Using Nonprotected Alkynyl Anchoring Groups}},
    year = {2016},
    journal = {Journal of the American Chemical Society},
    author = {Olavarria-Contreras, Ignacio José and Perrin, Mickael L. and Chen, Zhi and Klyatskaya, Svetlana and Ruben, Mario and Van Der Zant, Herre S.J.},
    number = {27},
    month = {7},
    pages = {8465--8469},
    volume = {138},
    publisher = {American Chemical Society},
    url = {/doi/pdf/10.1021/jacs.6b03383?ref=article_openPDF},
    doi = {10.1021/JACS.6B03383},
    issn = {15205126}
}

@article{Goswami2021DecisionMemristor,
    title = {{Decision trees within a molecular memristor}},
    year = {2021},
    journal = {Nature 2021 597:7874},
    author = {Goswami, Sreetosh and Pramanick, Rajib and Patra, Abhijeet and Rath, Santi Prasad and Foltin, Martin and Ariando, A. and Thompson, Damien and Venkatesan, T. and Goswami, Sreebrata and Williams, R. Stanley},
    number = {7874},
    month = {9},
    pages = {51--56},
    volume = {597},
    publisher = {Nature Publishing Group},
    url = {https://www.nature.com/articles/s41586-021-03748-0},
    doi = {10.1038/s41586-021-03748-0},
    issn = {1476-4687},
    pmid = {34471273}
}

@article{Venkataraman2006DependenceConformation,
    title = {{Dependence of single-molecule junction conductance on molecular conformation}},
    year = {2006},
    journal = {Nature 2006 442:7105},
    author = {Venkataraman, Latha and Klare, Jennifer E. and Nuckolls, Colin and Hybertsen, Mark S. and Steigerwald, Michael L.},
    number = {7105},
    month = {8},
    pages = {904--907},
    volume = {442},
    publisher = {Nature Publishing Group},
    url = {https://www.nature.com/articles/nature05037},
    doi = {10.1038/nature05037},
    issn = {1476-4687}
}

@article{Tanimoto2016DipoleJunctions,
    title = {{Dipole effects on the formation of molecular junctions}},
    year = {2016},
    journal = {Nanoscale Horizons},
    author = {Tanimoto, Sachie and Tsutsui, Makusu and Yokota, Kazumichi and Taniguchi, Masateru},
    number = {5},
    month = {8},
    pages = {399--406},
    volume = {1},
    publisher = {The Royal Society of Chemistry},
    url = {https://pubs.rsc.org/en/content/articlehtml/2016/nh/c6nh00088f https://pubs.rsc.org/en/content/articlelanding/2016/nh/c6nh00088f},
    doi = {10.1039/C6NH00088F},
    issn = {2055-6764}
}

@article{Kumar2022DynamicalComputing,
    title = {{Dynamical memristors for higher-complexity neuromorphic computing}},
    year = {2022},
    journal = {Nature Reviews Materials 2022 7:7},
    author = {Kumar, Suhas and Wang, Xinxin and Strachan, John Paul and Yang, Yuchao and Lu, Wei D.},
    number = {7},
    month = {4},
    pages = {575--591},
    volume = {7},
    publisher = {Nature Publishing Group},
    url = {https://www.nature.com/articles/s41578-022-00434-z},
    isbn = {0123456789},
    doi = {10.1038/s41578-022-00434-z},
    issn = {2058-8437}
}

@article{Li2015ElectricJunctions,
    title = {{Electric Field Breakdown in Single Molecule Junctions}},
    year = {2015},
    journal = {Journal of the American Chemical Society},
    author = {Li, Haixing and Su, Timothy A. and Zhang, Vivian and Steigerwald, Michael L. and Nuckolls, Colin and Venkataraman, Latha},
    number = {15},
    month = {4},
    pages = {5028--5033},
    volume = {137},
    publisher = {American Chemical Society},
    url = {/doi/pdf/10.1021/ja512523r?ref=article_openPDF},
    doi = {10.1021/JA512523R},
    issn = {15205126}
}

@article{Tang2020ElectricJunctions,
    title = {{Electric Field-Induced Assembly in Single-Stacking Terphenyl Junctions}},
    year = {2020},
    journal = {Journal of the American Chemical Society},
    author = {Tang, Yongxiang and Zhou, Yu and Zhou, Dahai and Chen, Yaorong and Xiao, Zongyuan and Shi, Jia and Liu, Junyang and Hong, Wenjing},
    number = {45},
    month = {11},
    pages = {19101--19109},
    volume = {142},
    publisher = {American Chemical Society},
    url = {/doi/pdf/10.1021/jacs.0c07348?ref=article_openPDF},
    doi = {10.1021/JACS.0C07348},
    issn = {15205126},
    pmid = {33135882}
}

@article{Alemani2006ElectricSTM,
    title = {{Electric field-induced isomerization of azobenzene by STM}},
    year = {2006},
    journal = {Journal of the American Chemical Society},
    author = {Alemani, Micol and Peters, Maike V. and Hecht, Stefan and Rieder, Karl Heinz and Moresco, Francesca and Grill, Leonhard},
    number = {45},
    month = {11},
    pages = {14446--14447},
    volume = {128},
    publisher = { American Chemical Society },
    url = {/doi/pdf/10.1021/ja065449s?ref=article_openPDF},
    doi = {10.1021/JA065449S},
    issn = {00027863}
}

@article{Gao2021ElectricJunctions,
    title = {{Electric field-induced switching among multiple conductance pathways in single-molecule junctions}},
    year = {2021},
    journal = {Chemical Communications},
    author = {Gao, Tengyang and Pan, Zhichao and Cai, Zhuanyun and Zheng, Jueting and Tang, Chun and Yuan, Saisai and Zhao, Shi qiang and Bai, Hua and Yang, Yang and Shi, Jia and Xiao, Zongyuan and Liu, Junyang and Hong, Wenjing},
    number = {58},
    month = {7},
    pages = {7160--7163},
    volume = {57},
    publisher = {The Royal Society of Chemistry},
    url = {https://pubs.rsc.org/en/content/articlehtml/2021/cc/d1cc02111g https://pubs.rsc.org/en/content/articlelanding/2021/cc/d1cc02111g},
    doi = {10.1039/D1CC02111G},
    issn = {1364-548X},
    pmid = {34184023}
}

@article{Olavarria-Contreras2018Electric-fieldCompounds,
    title = {{Electric-field induced bistability in single-molecule conductance measurements for boron coordinated curcuminoid compounds}},
    year = {2018},
    journal = {Chemical Science},
    author = {Olavarr{\'{i}}a-Contreras, Ignacio José and Etcheverry-Berr{\'{i}}os, Alvaro and Qian, Wenjie and Guti{\'{e}}rrez-Cer{\'{o}}n, Cristian and Campos-Olgu{\'{i}}n, Aldo and Sa{\~{n}}udo, E. Carolina and Duli{\'{c}}, Diana and Ruiz, Eliseo and Aliaga-Alcalde, Núria and Soler, Monica and Van Der Zant, Herre S.J.},
    number = {34},
    month = {8},
    pages = {6988--6996},
    volume = {9},
    publisher = {The Royal Society of Chemistry},
    url = {https://pubs.rsc.org/en/content/articlehtml/2018/sc/c8sc02337a https://pubs.rsc.org/en/content/articlelanding/2018/sc/c8sc02337a},
    doi = {10.1039/C8SC02337A},
    issn = {2041-6539}
}

@article{Han2020Electric-field-drivenJunctions,
    title = {{Electric-field-driven dual-functional molecular switches in tunnel junctions}},
    year = {2020},
    journal = {Nature Materials},
    author = {Han, Yingmei and Nickle, Cameron and Zhang, Ziyu and Astier, Hippolyte P.A.G. and Duffin, Thorin J. and Qi, Dongchen and Wang, Zhe and del Barco, Enrique and Thompson, Damien and Nijhuis, Christian A.},
    number = {8},
    month = {8},
    pages = {843--848},
    volume = {19},
    publisher = {Nature Research},
    url = {https://www.nature.com/articles/s41563-020-0697-5},
    doi = {10.1038/S41563-020-0697-5},
    issn = {14764660},
    pmid = {32483243}
}

@article{Tang2020Electric-Field-InducedJunctions,
    title = {{Electric-Field-Induced Connectivity Switching in Single-Molecule Junctions}},
    year = {2020},
    journal = {iScience},
    author = {Tang, Chun and Zheng, Jueting and Ye, Yiling and Liu, Junyang and Chen, Lijue and Yan, Zhewei and Chen, Zhixin and Chen, Lichuan and Huang, Xiaoyan and Bai, Jie and Chen, Zhaobin and Shi, Jia and Xia, Haiping and Hong, Wenjing},
    number = {1},
    month = {1},
    pages = {100770},
    volume = {23},
    publisher = {Elsevier},
    url = {https://www.sciencedirect.com/science/article/pii/S2589004219305152?via%3Dihub},
    doi = {10.1016/J.ISCI.2019.100770},
    issn = {2589-0042}
}

@article{Gantenbein2019ExploringDerivatives,
    title = {{Exploring antiaromaticity in single-molecule junctions formed from biphenylene derivatives}},
    year = {2019},
    journal = {Nanoscale},
    author = {Gantenbein, Markus and Li, Xiaohui and Sangtarash, Sara and Bai, Jie and Olsen, Gunnar and Alqorashi, Afaf and Hong, Wenjing and Lambert, Colin J. and Bryce, Martin R.},
    number = {43},
    month = {11},
    pages = {20659--20666},
    volume = {11},
    publisher = {The Royal Society of Chemistry},
    url = {https://pubs.rsc.org/en/content/articlehtml/2019/nr/c9nr05375a https://pubs.rsc.org/en/content/articlelanding/2019/nr/c9nr05375a},
    doi = {10.1039/C9NR05375A},
    issn = {2040-3372},
    pmid = {31641715}
}

@article{Ielmini2018In-memoryDevices,
    title = {{In-memory computing with resistive switching devices}},
    year = {2018},
    journal = {Nature Electronics 2018 1:6},
    author = {Ielmini, Daniele and Wong, H. S.Philip},
    number = {6},
    month = {6},
    pages = {333--343},
    volume = {1},
    publisher = {Nature Publishing Group},
    url = {https://www.nature.com/articles/s41928-018-0092-2},
    doi = {10.1038/s41928-018-0092-2},
    issn = {2520-1131}
}

@article{Leary2015IncorporatingGroup,
    title = {{Incorporating single molecules into electrical circuits. The role of the chemical anchoring group}},
    year = {2015},
    journal = {Chemical Society Reviews},
    author = {Leary, Edmund and La Rosa, Andrea and Gonz{\'{a}}lez, M. Teresa and Rubio-Bollinger, Gabino and Agra{\"{i}}t, Nicolás and Mart{\'{i}}n, Nazario},
    number = {4},
    month = {2},
    pages = {920--942},
    volume = {44},
    publisher = {The Royal Society of Chemistry},
    url = {https://pubs.rsc.org/en/content/articlehtml/2015/cs/c4cs00264d https://pubs.rsc.org/en/content/articlelanding/2015/cs/c4cs00264d},
    doi = {10.1039/C4CS00264D},
    issn = {1460-4744}
}

@article{SebastiaanvanderPoel2025InterferingJunctions,
    title = {{Interfering Electron Paths In Single-Molecule Junctions}},
    year = {2025},
    journal = {Delft University of Technology, PhD thesis},
    author = {{Sebastiaan van der Poel}},
    url = {http://doi.org/10.4121/638085f9-390b-438e-9b8e-c6a2d2310224.},
    isbn = {9789463847704},
    doi = {10.4121/638085f9-390b-438e-9b8e-c6a2d2310224}
}

@article{Lee2010IntramolecularBinding,
    title = {{Intramolecular Hydrogen Bonds Preorganize an Aryl-triazole Receptor into a Crescent for Chloride Binding}},
    year = {2010},
    journal = {Organic Letters},
    author = {Lee, Semin and Hua, Yuran and Park, Hyunsoo and Flood, Amar H.},
    number = {9},
    month = {5},
    pages = {2100--2102},
    volume = {12},
    publisher = { American Chemical Society},
    url = {/doi/pdf/10.1021/ol1005856?ref=article_openPDF},
    isbn = {1325313262},
    doi = {10.1021/OL1005856},
    issn = {15237060}
}

@article{Bandyopadhyay2010MassivelyLayer,
    title = {{Massively parallel computing on an organic molecular layer}},
    year = {2010},
    journal = {Nature Physics 2010 6:5},
    author = {Bandyopadhyay, Anirban and Pati, Ranjit and Sahu, Satyajit and Peper, Ferdinand and Fujita, Daisuke},
    number = {5},
    month = {4},
    pages = {369--375},
    volume = {6},
    publisher = {Nature Publishing Group},
    url = {https://www.nature.com/articles/nphys1636},
    doi = {10.1038/nphys1636},
    issn = {1745-2481}
}

@article{Haiss2004MeasurementWires,
    title = {{Measurement of single molecule conductivity using the spontaneous formation of molecular wires}},
    year = {2004},
    journal = {Physical Chemistry Chemical Physics},
    author = {Haiss, Wolfgang and Nichols, Richard J. and Van Zalinge, Harm and Higgins, Simon J. and Bethell, Donald and Schiffrin, David J.},
    number = {17},
    month = {8},
    pages = {4330--4337},
    volume = {6},
    publisher = {The Royal Society of Chemistry},
    url = {https://pubs.rsc.org/en/content/articlehtml/2004/cp/b404929b https://pubs.rsc.org/en/content/articlelanding/2004/cp/b404929b},
    doi = {10.1039/B404929B},
    issn = {1463-9084}
}

@article{Frisenda2016MechanicallyDimers,
    title = {{Mechanically controlled quantum interference in individual {$\pi$}-stacked dimers}},
    year = {2016},
    journal = {Nature Chemistry 2016 8:12},
    author = {Frisenda, Riccardo and Janssen, Vera A.E.C. and Grozema, Ferdinand C. and Van Der Zant, Herre S.J. and Renaud, Nicolas},
    number = {12},
    month = {8},
    pages = {1099--1104},
    volume = {8},
    publisher = {Nature Publishing Group},
    url = {https://www.nature.com/articles/nchem.2588},
    doi = {10.1038/nchem.2588},
    issn = {1755-4349}
}

@article{Qiu2004MechanismsMolecule,
    title = {{Mechanisms of Reversible Conformational Transitions in a Single Molecule}},
    year = {2004},
    journal = {Physical Review Letters},
    author = {Qiu, X. H. and Nazin, G. V. and Ho, W.},
    number = {19},
    month = {11},
    pages = {196806},
    volume = {93},
    publisher = {American Physical Society},
    url = {https://journals.aps.org/prl/abstract/10.1103/PhysRevLett.93.196806},
    doi = {10.1103/PhysRevLett.93.196806},
    issn = {00319007}
}

@article{Chua1971Memristor-TheElement,
    title = {{Memristor-The Missing Circuit Element}},
    year = {1971},
    journal = {EEE Transactions on Circuit Theory},
    author = {Chua, Leon O.},
    number = {5},
    pages = {507},
    volume = {18}
}

@article{He2005Metal-freeDevice,
    title = {{Metal-free silicon – molecule – nanotube testbed and memory device}},
    year = {2005},
    journal = {Nature Materials 2006 5:1},
    author = {He, Jianli and Chen, Bo and Flatt, Austen K. and Stephenson, Jason J. and Doyle, Condell D. and Tour, James M.},
    number = {1},
    month = {12},
    pages = {63--68},
    volume = {5},
    publisher = {Nature Publishing Group},
    url = {https://www.nature.com/articles/nmat1526},
    doi = {10.1038/nmat1526},
    issn = {1476-4660},
    pmid = {16327789}
}

@article{Gorenskaia2024MethodsBehaviour,
    title = {{Methods for the analysis, interpretation, and prediction of single-molecule junction conductance behaviour}},
    year = {2024},
    journal = {Chemical Science},
    author = {Gorenskaia, Elena and Low, Paul J.},
    number = {25},
    month = {5},
    pages = {9510},
    volume = {15},
    publisher = {Royal Society of Chemistry},
    url = {https://pmc.ncbi.nlm.nih.gov/articles/PMC11206205/},
    doi = {10.1039/D4SC00488D},
    issn = {20416539},
    pmid = {38939131}
}

@book{Cuevas2010MolecularExperiment,
    title = {{Molecular electronics : an introduction to theory and experiment}},
    year = {2010},
    author = {Cuevas, Juan Carlos. and Scheer, Elke.},
    edition = {},
    pages = {724},
    publisher = {World Scientific},
    url = {https://books.google.com/books/about/Molecular_Electronics.html?hl=it&id=VtZpDQAAQBAJ},
    isbn = {9814282588}
}

@article{VonHippel1956MolecularEngineering,
    title = {{Molecular Engineering}},
    year = {1956},
    journal = {Science},
    author = {Von Hippel, A.},
    number = {3191},
    month = {2},
    pages = {315--317},
    volume = {123},
    publisher = {American Association for the Advancement of Science},
    url = {/doi/pdf/10.1126/science.123.3191.315?download=true},
    doi = {10.1126/SCIENCE.123.3191.315},
    issn = {00368075},
    pmid = {17774519}
}

@article{Troisi2006Molecularmolecular,
    title = {{Molecular signatures in the transport properties of molecular wire junctions: What makes a junction "molecular"?}},
    year = {2006},
    journal = {Small},
    author = {Troisi, Alessandro and Ratner, Mark A.},
    number = {2},
    month = {2},
    pages = {172--181},
    volume = {2},
    doi = {10.1002/SMLL.200500201},
    issn = {16136810},
    pmid = {17193017}
}

@article{Blum2005MolecularlySwitching,
    title = {{Molecularly inherent voltage-controlled conductance switching}},
    year = {2005},
    journal = {Nature Materials 2005 4:2},
    author = {Blum, Amy Szuchmacher and Kushmerick, James G. and Long, David P. and Patterson, Charles H. and Yang, John C. and Henderson, Jay C. and Yao, Yuxing and Tour, James M. and Shashidhar, Ranganathan and Ratna, Banahalli R.},
    number = {2},
    month = {1},
    pages = {167--172},
    volume = {4},
    publisher = {Nature Publishing Group},
    url = {https://www.nature.com/articles/nmat1309},
    doi = {10.1038/nmat1309},
    issn = {1476-4660}
}

@article{Yan2010Molecularly-mediatedStructures,
    title = {{Molecularly-mediated assembly of gold nanoparticles with interparticle rigid, conjugated and shaped aryl ethynyl structures}},
    year = {2010},
    journal = {Chemical Communications},
    author = {Yan, Hong and Lim, Stephanie I. and Zhang, Ying Jun and Chen, Qiang and Mott, Derrick and Wu, Wei Tai and An, De Lie and Zhou, Shuiqin and Zhong, Chuan Jian},
    number = {13},
    month = {3},
    pages = {2218--2220},
    volume = {46},
    publisher = {The Royal Society of Chemistry},
    url = {https://pubs.rsc.org/en/content/articlehtml/2010/cc/b927496k https://pubs.rsc.org/en/content/articlelanding/2010/cc/b927496k},
    doi = {10.1039/B927496K},
    issn = {1364-548X},
    pmid = {20234911}
}

@article{Stewart2003Molecule-IndependentDevices,
    title = {{Molecule-Independent Electrical Switching in Pt/Organic Monolayer/Ti Devices}},
    year = {2003},
    journal = {Nano Letters},
    author = {Stewart, D. R. and Ohlberg, D. A.A. and Beck, P. A. and Chen, Y. and Williams, R. Stanley and Jeppesen, J. O. and Nielsen, K. A. and Stoddart, J. Fraser},
    number = {1},
    month = {1},
    pages = {133--136},
    volume = {4},
    publisher = { American Chemical Society },
    url = {/doi/pdf/10.1021/nl034795u?ref=article_openPDF},
    doi = {10.1021/NL034795U},
    issn = {15306984}
}

@article{Jakobsson2007OnDevices,
    title = {{On the switching mechanism in Rose Bengal-based memory devices}},
    year = {2007},
    journal = {Organic Electronics},
    author = {Jakobsson, Fredrik L.E. and Crispin, Xavier and C{\"{o}}lle, Michael and B{\"{u}}chel, Michael and de Leeuw, Dago M. and Berggren, Magnus},
    number = {5},
    month = {10},
    pages = {559--565},
    volume = {8},
    publisher = {North-Holland},
    url = {https://www.sciencedirect.com/science/article/pii/S1566119907000493},
    doi = {10.1016/J.ORGEL.2007.04.002},
    issn = {1566-1199}
}

@article{Frisenda2015OPE3:Transport,
    title = {{OPE3: A Model System For Single-Molecule Transport}},
    year = {2015},
    journal = {Delft University of Technology, PhD thesis},
    author = {Frisenda, Riccardo},
    publisher = {},
    url = {http://repository.tudelft.nl/.},
    isbn = {9789085932406}
}

@article{Prime2009OverviewDevices,
    title = {{Overview of organic memory devices}},
    year = {2009},
    journal = {Philosophical Transactions of the Royal Society A: Mathematical, Physical and Engineering Sciences},
    author = {Prime, D. and Paul, S.},
    number = {1905},
    month = {10},
    pages = {4141--4157},
    volume = {367},
    publisher = {The Royal Society},
    url = {https://scholar.google.com/scholar_url?url=https://royalsocietypublishing.org/doi/pdf/10.1098/rsta.2009.0165&hl=it&sa=T&oi=ucasa&ct=ufr&ei=8XRJafidGIDO6rQP-qaaoAo&scisig=ALhkC2RGG9tNoww3F8KAHQEA0FAg},
    doi = {10.1098/RSTA.2009.0165},
    issn = {1364-503X}
}

@article{Frisenda2018QuantumStudy,
    title = {{Quantum Transport through a Single Conjugated Rigid Molecule, a Mechanical Break Junction Study}},
    year = {2018},
    journal = {Accounts of Chemical Research},
    author = {Frisenda, Riccardo and Stefani, Davide and Van Der Zant, Herre S.J.},
    number = {6},
    month = {6},
    pages = {1359--1367},
    volume = {51},
    publisher = {American Chemical Society},
    doi = {10.1021/acs.accounts.7b00493},
    issn = {15204898},
    pmid = {29862817}
}

@article{Xie2024RegularlyEffects,
    title = {{Regularly Tuning Quantum Interference in Single-Molecule Junctions through Systematic Substitution of Side Groups with Varied Electron Effects†}},
    year = {2024},
    journal = {Chinese Journal of Chemistry},
    author = {Xie, Xianjing and Zhang, Yirong and Zhang, Junrui and Cui, Xingyuan and Liu, Wei and Liu, Xunshan},
    number = {11},
    month = {6},
    pages = {1217--1222},
    volume = {42},
    publisher = {Shanghai Institute of Organic Chemistry},
    url = {/doi/pdf/10.1002/cjoc.202300706 https://onlinelibrary.wiley.com/doi/abs/10.1002/cjoc.202300706 https://onlinelibrary.wiley.com/doi/10.1002/cjoc.202300706},
    doi = {10.1002/CJOC.202300706},
    issn = {16147065}
}

@article{Batista2007ResistiveNanocontacts,
    title = {{Resistive and rectifying effects of pulling gold atoms at thiol-gold nanocontacts}},
    year = {2007},
    journal = {Physical Review B},
    author = {Batista, Ronaldo J.C. and Ordej{\'{o}}n, Pablo and Chacham, Helio and Artacho, Emilio},
    number = {4},
    month = {1},
    pages = {041402},
    volume = {75},
    publisher = {American Physical Society},
    url = {https://journals.aps.org/prb/abstract/10.1103/PhysRevB.75.041402},
    doi = {10.1103/PhysRevB.75.041402},
    issn = {10980121}
}

@article{Chen2025ResponsiveDiversification,
    title = {{Responsive Molecules for Organic Neuromorphic Devices: Harnessing Memory Diversification}},
    year = {2025},
    journal = {Advanced Materials},
    author = {Chen, Yusheng and Han, Bin and Gobbi, Marco and Hou, Lili and Samor{\`{i}}, Paolo},
    number = {19},
    month = {5},
    pages = {2418281},
    volume = {37},
    publisher = {John Wiley and Sons Inc},
    url = {10.1002/adma.202418281 https://onlinelibrary.wiley.com/doi/abs/10.1002/adma.202418281 https://advanced.onlinelibrary.wiley.com/doi/10.1002/adma.202418281},
    doi = {10.1002/ADMA.202418281},
    issn = {15214095},
    pmid = {40135253}
}

@article{Lortscher2006ReversibleJunction,
    title = {{Reversible and controllable switching of a single-molecule junction}},
    year = {2006},
    journal = {Small},
    author = {L{\"{o}}rtscher, Emanuel and Ciszek, Jacob W. and Tour, James and Riel, Heike},
    number = {8-9},
    month = {8},
    pages = {973--977},
    volume = {2},
    doi = {10.1002/SMLL.200600101},
    issn = {16136810},
    pmid = {17193152}
}

@article{Goswami2017RobustAzoaromatics,
    title = {{Robust resistive memory devices using solution-processable metal-coordinated azo aromatics}},
    year = {2017},
    journal = {Nature Materials 2017 16:12},
    author = {Goswami, Sreetosh and Matula, Adam J. and Rath, Santi P. and Hedstr{\"{o}}m, Svante and Saha, Surajit and Annamalai, Meenakshi and Sengupta, Debabrata and Patra, Abhijeet and Ghosh, Siddhartha and Jani, Hariom and Sarkar, Soumya and Motapothula, Mallikarjuna Rao and Nijhuis, Christian A. and Martin, Jens and Goswami, Sreebrata and Batista, Victor S. and Venkatesan, T.},
    number = {12},
    month = {10},
    pages = {1216--1224},
    volume = {16},
    publisher = {Nature Publishing Group},
    url = {https://www.nature.com/articles/nmat5009},
    doi = {10.1038/nmat5009},
    issn = {1476-4660},
    pmid = {29058729}
}

@article{Ezenwafor2025SupramolecularPrinciples,
    title = {{Supramolecular Protection of Carboxylic Acids via Hydrogen Bonding: Selectivity, Reversibility, and Design Principles}},
    year = {2025},
    journal = {Chemistry - A European Journal},
    author = {Ezenwafor, Oluebube F. and Liu, Hao and Shimizu, Ken D.},
    number = {55},
    month = {10},
    pages = {e02034},
    volume = {31},
    publisher = {John Wiley and Sons Inc},
    url = {/doi/pdf/10.1002/chem.202502034 https://onlinelibrary.wiley.com/doi/abs/10.1002/chem.202502034 https://chemistry-europe.onlinelibrary.wiley.com/doi/10.1002/chem.202502034},
    doi = {10.1002/CHEM.202502034;ISSUE:ISSUE:DOI},
    issn = {15213765},
    pmid = {40727952}
}

@article{Ornago2022SwitchingReconfiguration,
    title = {{Switching in Nanoscale Molecular Junctions due to Contact Reconfiguration}},
    year = {2022},
    journal = {The Journal of Physical Chemistry C},
    author = {Ornago, Luca and Kamer, Jerry and El Abbassi, Maria and Grozema, Ferdinand C. and Van Der Zant, Herre S.J.},
    number = {46},
    month = {11},
    pages = {19843--19848},
    volume = {126},
    publisher = {American Chemical Society},
    url = {/doi/pdf/10.1021/acs.jpcc.2c04370?ref=article_openPDF},
    doi = {10.1021/ACS.JPCC.2C04370},
    issn = {19327455}
}

@article{Miguel2015TowardDerivatives,
    title = {{Toward Multiple Conductance Pathways with Heterocycle-Based Oligo (phenyleneethynylene) Derivatives}},
    year = {2015},
    journal = {Journal of the American Chemical Society},
    author = {Miguel, Delia and {\'{A}}lvarez De Cienfuegos, Luis and Mart{\'{i}}n-Lasanta, Ana and Morcillo, Sara P. and Zotti, Linda A. and Leary, Edmund and B{\"{u}}rkle, Marius and Asai, Yoshihiro and Jurado, Rocío and C{\'{a}}rdenas, Diego J. and Rubio-Bollinger, Gabino and Agra{\"{i}}t, Nicolás and Cuerva, Juan M. and Gonz{\'{a}}lez, M. Teresa},
    number = {43},
    month = {11},
    pages = {13818--13826},
    volume = {137},
    publisher = {American Chemical Society},
    url = {/doi/pdf/10.1021/jacs.5b05637?ref=article_openPDF},
    doi = {10.1021/JACS.5B05637},
    issn = {15205126}
}

@article{Jiang2019TurningConductance,
    title = {{Turning the Tap: Conformational Control of Quantum Interference to Modulate Single-Molecule Conductance}},
    year = {2019},
    journal = {Angewandte Chemie - International Edition},
    author = {Jiang, Feng and Trupp, Douglas I. and Algethami, Norah and Zheng, Haining and He, Wenxiang and Alqorashi, Afaf and Zhu, Chenxu and Tang, Chun and Li, Ruihao and Liu, Junyang and Sadeghi, Hatef and Shi, Jia and Davidson, Ross and Korb, Marcus and Sobolev, Alexandre N. and Naher, Masnun and Sangtarash, Sara and Low, Paul J. and Hong, Wenjing and Lambert, Colin J.},
    number = {52},
    month = {12},
    pages = {18987--18993},
    volume = {58},
    publisher = {Wiley-VCH Verlag},
    url = {/doi/pdf/10.1002/anie.201909461 https://onlinelibrary.wiley.com/doi/abs/10.1002/anie.201909461 https://onlinelibrary.wiley.com/doi/10.1002/anie.201909461},
    doi = {10.1002/ANIE.201909461},
    issn = {15213773},
    pmid = {31617293}
}

@article{Soni2020UnderstandingJunctions,
    title = {{Understanding the Role of Parallel Pathways via In-Situ Switching of Quantum Interference in Molecular Tunneling Junctions}},
    year = {2020},
    journal = {Angewandte Chemie - International Edition},
    author = {Soni, Saurabh and Ye, Gang and Zheng, Jueting and Zhang, Yanxi and Asyuda, Andika and Zharnikov, Michael and Hong, Wenjing and Chiechi, Ryan C.},
    number = {34},
    month = {8},
    pages = {14308--14312},
    volume = {59},
    publisher = {Wiley-VCH Verlag},
    url = {/doi/pdf/10.1002/anie.202005047 https://onlinelibrary.wiley.com/doi/abs/10.1002/anie.202005047 https://onlinelibrary.wiley.com/doi/10.1002/anie.202005047},
    doi = {10.1002/ANIE.202005047},
    issn = {15213773},
    pmid = {32469444}
}

@article{Wang2012Voltage-dependentJunction,
    title = {{Voltage-dependent conductance states of a single-molecule junction}},
    year = {2012},
    journal = {Journal of Physics: Condensed Matter},
    author = {Wang, Y. F. and N{\'{e}}el, N. and Kr{\"{o}}ger, J. and V{\'{a}}zquez, H. and Brandbyge, M. and Wang, B. and Berndt, R.},
    number = {39},
    month = {9},
    pages = {394012},
    volume = {24},
    publisher = {IOP Publishing},
    url = {https://iopscience.iop.org/article/10.1088/0953-8984/24/39/394012 https://iopscience.iop.org/article/10.1088/0953-8984/24/39/394012/meta},
    isbn = {131.180.33.35},
    doi = {10.1088/0953-8984/24/39/394012},
    issn = {09538984}
}

@article{Bandyopadhyay2006WritingMicroscope,
    title = {{Writing and erasing information in multilevel logic systems of a single molecule using scanning tunneling microscope}},
    year = {2006},
    journal = {Applied Physics Letters},
    author = {Bandyopadhyay, Anirban and Miki, K. and Wakayama, Y.},
    number = {24},
    month = {12},
    pages = {243506},
    volume = {89},
    publisher = {AIP Publishing},
    url = {/aip/apl/article/89/24/243506/508652/Writing-and-erasing-information-in-multilevel},
    doi = {10.1063/1.2402895},
    issn = {00036951}
}
\newpage
\begin{figure}[ht]
\centering
\includegraphics[width=\linewidth]{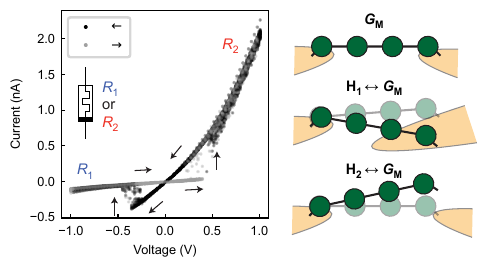}
\caption{TOC Graphic}
\label{toc}
\end{figure}
\end{document}

\renewcommand{\thepage}{S\arabic{page}}

\maketitle

\vspace{0.5cm}

\tableofcontents
\newpage
\clearpage
\appendix % optional; for section letters; not required for S-numbering

% --- begin: S-numbering only within SI ---
\newcommand{\beginsupplement}{
  \setcounter{figure}{0}
  \renewcommand{\thefigure}{S\arabic{figure}}
  \setcounter{table}{0}
  \renewcommand{\thetable}{S\arabic{table}}
  \setcounter{equation}{0}
  \renewcommand{\theequation}{S\arabic{equation}}
}
\beginsupplement
% --- end: S-numbering only within SI ---
\section{Device Fabrication}
The fabrication process used to make MCBJ devices is based on the one developed previously in the group
\cite{Perrin2015ChargeTheory, Martin2011AJunctions}.
\begin{itemize}
  \item \textbf{Substrate:} Phosphor bronze, $50\times50\times0.5\,\mathrm{mm}$ (flexible).
  \item \textbf{Insulating layer:} Polyimide (PI2610, HD Microsystems) deposited on the substrate. Prior to coating, apply adhesion promoter VM651 (HD Microsystems).
  \item \textbf{Nanowire patterning and deposition:} Electron-beam lithography (Raith EBPG5200 HS) followed by metal evaporation (Temescal FC-2000): $3\,\mathrm{nm}$ Ti / $80\,\mathrm{nm}$ Au, defining a $50\,\mathrm{nm}$-wide constriction at the center.
  \item \textbf{Suspension etching:} Oxygen reactive-ion etching (Sentech RIE) to remove $\sim 1\,\mu\mathrm{m}$ of polyimide and suspend the Au bridge of the MCBJ.
\end{itemize}

\section{Room-temperature transport measurements}
\begin{figure}[ht]
\centering
\includegraphics[width=\linewidth]{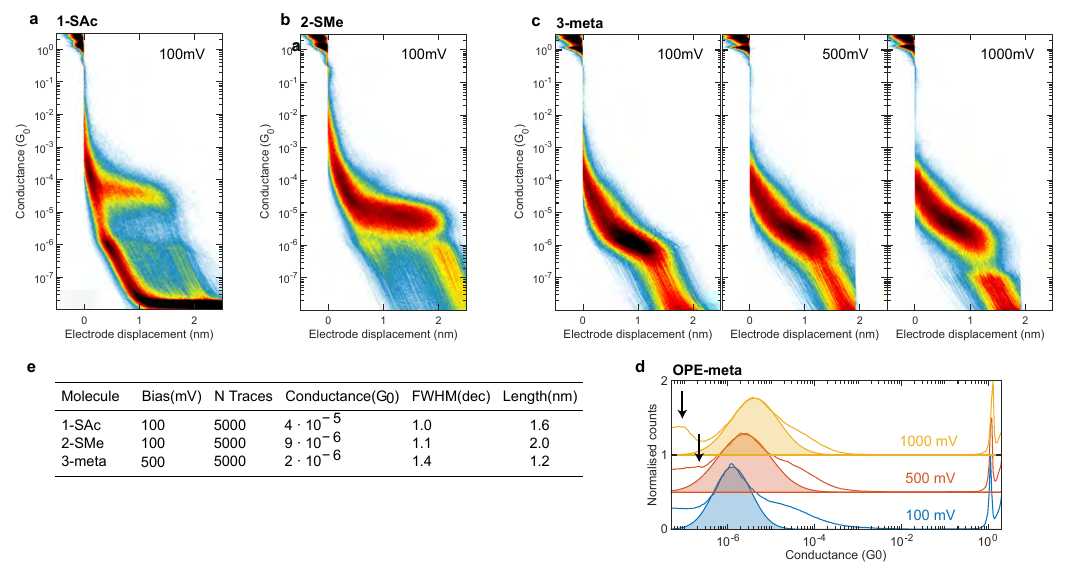}
\caption{2D conductance histograms and summary.
(a–b) Two-dimensional conductance histograms for \textbf{1–SAc} and \textbf{2–SMe}, each built from $5{,}000$ room-temperature breaking traces at $100\,\mathrm{mV}$.
(c) \textbf{3–meta} histograms measured at multiple biases.
(d) 1D histograms showing the artifact’s shift (black arrow) to lower conductance with increasing bias and the bias-dependent evolution of the molecular feature.
(e) Table summarizing key parameters and results of the room-temperature measurements.}
\label{roomT_appendix}
\end{figure}

\noindent Room-temperature measurements were performed in a custom-made mechanically controlled break-junction (MCBJ) setup, described in detail in \cite{Ornago2023ComplexityJunctions, SebastiaanvanderPoel2025InterferingJunctions, Martin2011AJunctions, Frisenda2015OPE3:Transport}. The contact was driven at $1\,\mathrm{nm\,s^{-1}}$ under an applied bias of $100\,\mathrm{mV}$ for \textbf{1–SAc} and \textbf{2–SMe}, while \textbf{3–meta} was measured at $100\,\mathrm{mV}$, $500\,\mathrm{mV}$, and $1\,\mathrm{V}$. The target molecules are dissolved in methanol (0.05-0.1 mmol). Before dropcasting, Au cleanliness was verified by acquiring $>500$ empty-junction traces per device. Horizontal displacement calibration is verified by checking the length of the $1\mathrm{G}_0$ plateau to be $\sim0.25\,\mathrm{nm}$.

To reduce noise and extend the current range, a PCB with a logarithmic amplifier was placed close to the junction \cite{Ornago2023ComplexityJunctions}. This introduces a reproducible artifact near $ \sim 1\cdot10^{-6}\,G_0 $ at $100\,\mathrm{mV}$ (corresponding to $\sim 0.8\, \mathrm{pA}$) \cite{Ornago2023ComplexityJunctions,SebastiaanvanderPoel2025InterferingJunctions}. Features at this conductance should not be interpreted as molecular. The artifact originates from the load-dependent settling time of the amplifier and shifts systematically with bias. The artifact moves to lower conductance at higher bias, enabling separation from genuine molecular signals.

Figure \ref{roomT_appendix}a–b shows 2D conductance histograms for \textbf{1–SAc} and \textbf{2–SMe} at $100\,\mathrm{mV}$. For \textbf{3–meta} (Fig. \ref{roomT_appendix}c), histograms at $100\,\mathrm{mV}$, $500\,\mathrm{mV}$, and $1\,\mathrm{V}$ were acquired because the molecular features overlap the amplifier artifact at $100\,\mathrm{mV}$. The histogram at $500\,\mathrm{mV}$ was selected for subsequent analysis as it separates the artifact from the molecular peak while remaining in a low-bias regime.

One-dimensional conductance histograms are obtained by projecting the 2D maps onto the conductance axis. A Gaussian fit to the main peak yields the mean conductance ($\mu$) and the full width at half maximum (FWHM), which define the conductance window of fully stretched junctions used in the main text (gray shaded area in Fig. 5b). 

Plateau-length histograms are constructed by computing, for each trace, the displacement span over which the conductance lies between a lower bound $\mu-\sigma$ (with $\mu$ and $\sigma$ the mean and standard deviation of the Gaussian fit) and an upper bound given by rupture threshold $3\cdot10^{-1}\,G_{0}$. These geometric lengths do not directly equal the molecular length, since Au snap-back and possible Au atom pull-out at the anchors are not yet corrected for. Notably, apparent plateau lengths with the $-\mathrm{SMe}$ anchor exceed those for $-\mathrm{SAc}$, suggesting greater junction stability for $-\mathrm{SAc}$ despite its weaker coupling.

\section{Low-temperature transport measurements}\label{section:low_t_measurements}
The MCBJ setup used for low-temperature measurements is similar to the room-temperature one, with two main differences: (i) a servomotor is used instead of a piezo actuator, providing a larger displacement window at the expense of precision and with increased break/make hysteresis; (ii) current readout is performed with external amplifiers: a home-made logarithmic amplifier is employed for breaking traces, albeit with a higher noise floor than the room-temperature setup (approximately $1\cdot10^{-6}\mathrm{G}_0$), while a low-noise linear amplifier is used for IV measurements to improve signal quality (FEMTO DLPCA-200).
A fast electronic switch selects the active measurement path. During breaking traces the applied bias is typically $200\,\mathrm{mV}$ and the vertical displacement speed is $1\,\mathrm{\mu m\,s^{-1}}$, corresponding, after mechanical attenuation, to a horizontal displacement of $\sim 5.5\,\mathrm{\textup{~\AA} \,s^{-1}}$.

\subsection{Measuring IVs}\label{sec:measuring_ivs}

\noindent A conductance window is defined where molecular signals are expected between the Au rupture and the amplifier noise floor (typically the window is from $10^{-2}$ to $10^{-6}\mathrm{G}_0$). Within this range, continuous motion is halted and stepped displacements of $0.4 \,\mathrm{\textup{~\AA}}$ (with intermediate $0.1 \,\mathrm{\textup{~\AA}}$ steps) are applied. After each step, an IV sweep is recorded over $-1\mathrm{V}$ to $+1\mathrm{V}$. The sweep starts and ends at $0\mathrm{V}$ and alternates the initial ramp between positive and negative bias. The positive ramp looks like $\sim0\mathrm{V}\rightarrow1\mathrm{V}$, $1\mathrm{V}\rightarrow-1\mathrm{V}$, $-1\mathrm{V}\rightarrow1\mathrm{V}$, and $1\mathrm{V}\rightarrow\sim0\mathrm{V}$. For clarity, only the two central segments are shown in the IV plots. 

\subsection{Empty and molecular traces} \label{sec:empty and molecular traces}

\noindent At low temperature, traces are classified as \emph{empty} or \emph{molecular} by the number of IVs recorded before reaching the noise floor. In particular, traces with fewer than 3 IVs (i.e., fewer than three motor steps) are considered empty.

\begin{figure}[ht]
\centering
\includegraphics[width=\linewidth]{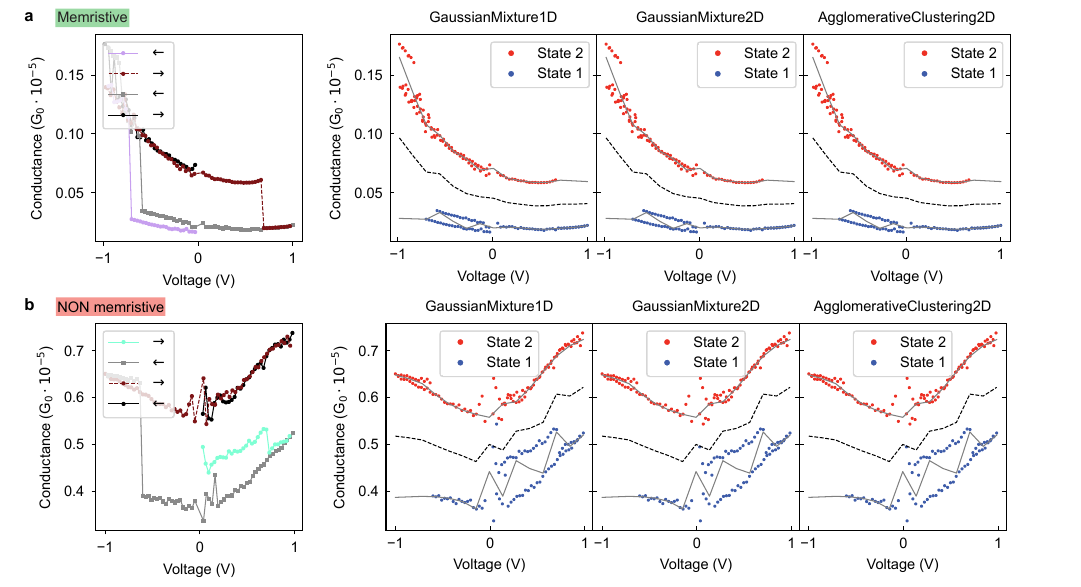}
\caption{Clustering and memristivity workflow. For each conductance-voltage (GV) curve, all four sweeping segments, including the ramping from and to zero bias, are considered. Three clustering methods are run on the datapoints with $k=2$ classes. The memristivity criteria are evaluated on each grouping, and a majority vote across the three outcomes determines the final memristive/non-memristive label. (a) Example GV trace classified as memristive. (b) Example GV trace classified as non-memristive, given that the hysteretic loop is not completed at positive bias.}
\label{clustering_appendix}
\end{figure}

\subsection{Memristive IV classification}\label{sec:meristive_iv_classification}

\noindent IVs are labeled as \emph{memristive} or \emph{non-memristive}, by the presence or absence, respectively, of non-volatile bistable switching under opposite sweep directions. An algorithmic procedure based on clustering of conductance values is used to divide points into two states, and then, a set of memristivity criteria is used for evaluation (\cref{clustering_appendix}). 

The analysis starts with the conductance-voltage (GV) representation of the data, including the ramps from and to zero bias.
Nonphysical conductance values (negative or zero, typically arising from imperfect zero-bias current offset correction) are replaced by the smallest positive conductance within the same dataset. 
The conductance values are then rescaled by a logarithmic feature scaling: 
$$
\tilde G_i = 10\,\log_{10}(G_i).
$$
This scaling, compared to a linear one, helps the clustering procedure to be invariant on the conductance amplitude, accounting for the wide range of values ($10^{-6}\mathrm{G}_0$ to $10^{-2}\mathrm{G}_0$). The multiplication factor $10$ further elongates the feature space to promote clustering into states with spread voltage distribution and clear conductance separation.

Three different clustering methods are used to classify the points into two groups: 2D Gaussianmixture model (G and V), 1D Gaussianmixture model (only G), and 2D agglomerative clustering (G and V). Benchmarks of the methods on a subset of data show small but recurring differences among methods. Therefore, we decided to proceed with all three classifiers and then make a decision based on majority voting. 

Because clustering will nominally produce two groups even without a genuine switch, additional quality criteria are enforced to identify true bistability and nonvolatility. The requirements are:

\begin{enumerate}\label{appendix_memristive_criteria}
    \item Both conductance states are sampled at low bias ($4\,\mathrm{mV}<|V|<12\,\mathrm{mV}$) in the complete sweep segments.
    \item The zero-bias conductance ratio satisfies $G_{2}/G_{1} \ge 2$. In addition, the interpolated state curves must remain pointwise separated by a ratio of $\ge 1.5$.
    \item State separation is significant: $3\sigma_1 + 3\sigma_2 < G_2 - G_1$, where $\sigma_{1,2}$ are the standard deviations and $G_{1,2}$ the mean conductances of the low and high states.
    \item Switching between the two states occurs at both polarities for $|V|>100\,\mathrm{mV}$, where the minimum bias is set to $100$mV to account for possible low-bias artifacts.
    \item Switching is observed on both up-sweep and down-sweep segments.
\end{enumerate}

To avoid false positives near the noise floor, an additional filter requires $G_2 \ge 4\cdot10^{-7}\,\mathrm{G}_0$.

\subsection{Stability of hold positions}\label{sec:stability_of_hold_positions}
\noindent A hold position is classified as \textit{stable} if at least 30 IV sweeps are acquired and at least $20\%$ of them are classified as memristive. Here, stability quantifies the reproducibility of switching at fixed displacement and therefore requires both that the molecule remains in the junction and that switching occurs sufficiently often. An IV collection is terminated when the average conductance drops below the noise floor, typically set to $10^{-7}\,\mathrm{G}_{0}$. Consequently, collections in which the molecule is lost during the bias sweeps contain fewer than 100 IVs. We require a minimum of $30$ sweeps to ensure adequate statistics at a given hold position. The $20\%$ threshold reflects two considerations: (i) memristive switching does not necessarily occur in every sweep, and (ii) our memristivity criteria are intentionally strict, so some switching events may not be classified as memristive. An overview of the workflow from measurement to raw data and (stable) IV collections is provided in \cref{measurement_protocol_appendix}.

\subsection{Search for memristive IVs}\label{sec:search_memristive_ivs}

\begin{figure}[H]
\centering
\includegraphics[width=\linewidth]{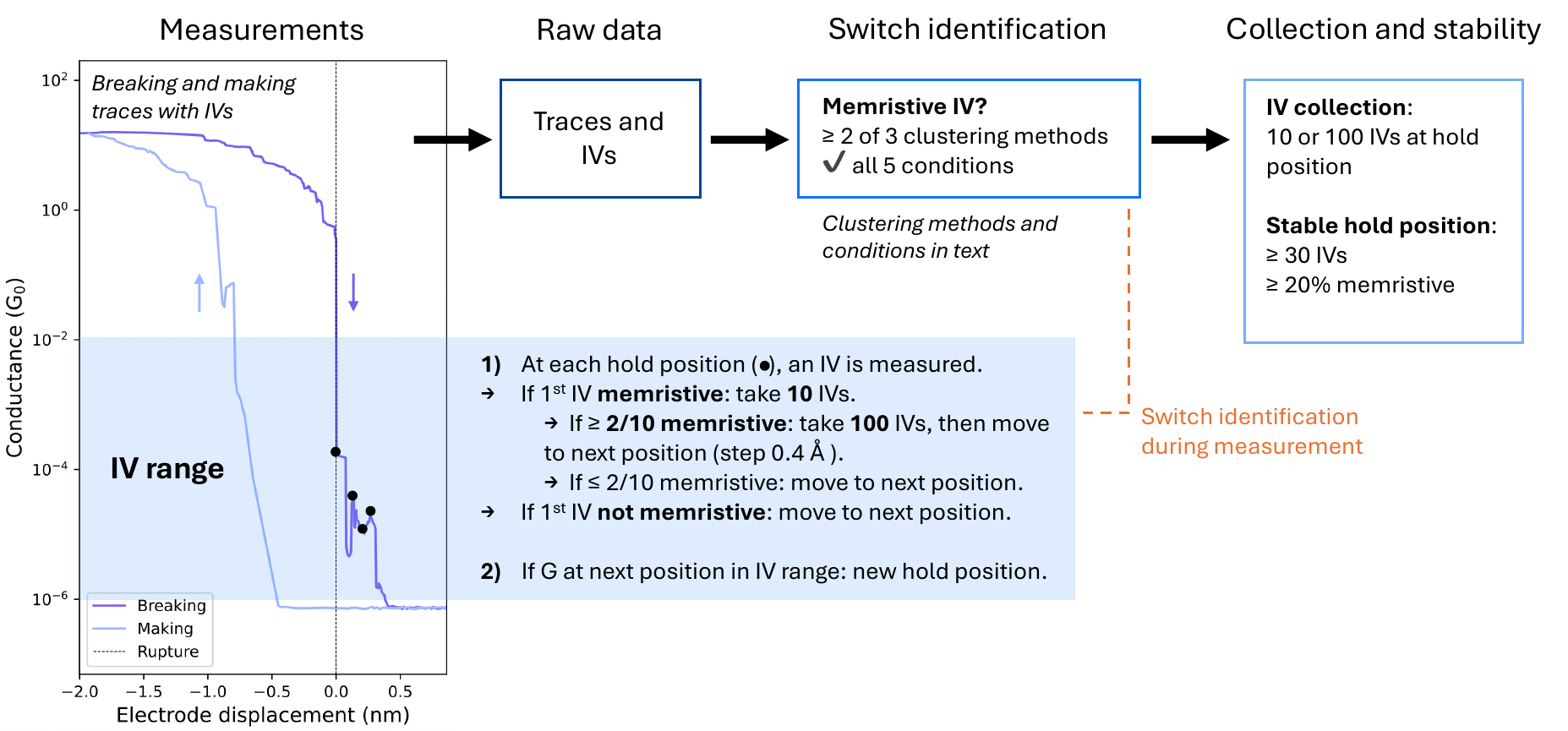}
\caption{Illustration of the measurement protocol and data processing to search for memristive hold positions and characterize their stability.}
\label{measurement_protocol_appendix}
\end{figure}

Figure \ref{measurement_protocol_appendix} summarizes the protocol used to search for and characterize nonvolatile, bistable memristive IVs. During each breaking trace, IVs are recorded when the conductance lies within the molecular window depicted as a blue shaded area in \cref{measurement_protocol_appendix}. IVs are recorded only on breaking traces (opening), and not on making traces (closing). For each sweep, the GV curve is clustered into two states, and the memristivity criteria are evaluated. If the hold position is classified as non-memristive, the motor is stepped to a new displacement. If the hold position is memristive, an additional 9 IVs are acquired. When at least 2 sweeps are identified as memristive (the switching can be reproduced), a further 90 IVs are recorded, yielding a set of 100 IVs for the same junction hold position (an \emph{IV collection}; examples in main text Fig. 3). As mentioned above, the IV collection is prematurely interrupted if the average IV conductance is below $10^{-7}\,\mathrm{G}_{0}$, indicating that the molecule is no longer in the junction. The motor is then stepped to obtain a new hold position, and the procedure is repeated while the conductance remains within the molecular range. 

\section{Memristive switchinng statistics}\label{sec:statistics_appendix}

\begin{figure}[H]
\centering
\includegraphics[width=\linewidth]{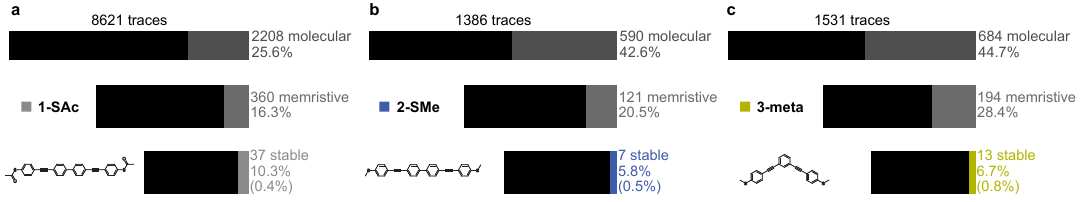}
\caption{For each molecule, the top bar indicates the percentage of molecular traces. Based on this subset, the fractions of memristive and stable traces are reported. Stability is shown relative to memristive traces and, in brackets, relative to all traces.}
\label{statistics_appendix}
\end{figure}

\begin{figure}[H]
\centering
\includegraphics[width=\linewidth]{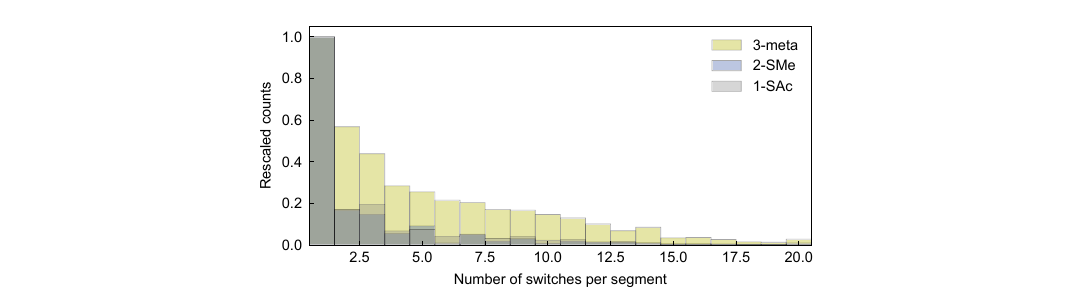}
\caption{  Number of switching events in each directional segment for the three target molecules.}
\label{number_switches}
\end{figure}

Figure~\ref{number_switches} reports the number of switching events observed within each unidirectional segment of memristive IV sweeps. For sweeps to the right we count only switching events at positive bias, whereas for sweeps to the left we count only events at negative bias. In the ideal case of purely field-driven memristive switching, each segment should contain a single switching event ($n=1$). 

Figure~\ref{number_switches} shows the resulting distributions for the three molecules. \textbf{1--SAc} and \textbf{2--SMe} display similar behavior, with the large majority of segments exhibiting $n=1$. In contrast, \textbf{3--meta} shows a broader distribution extending to higher switching counts. This trend supports the interpretation that switching in \textbf{1--SAc} and \textbf{2--SMe} is predominantly field-driven, whereas \textbf{3--meta} exhibits a larger contribution from current-driven, stochastic switching.
 
\section{Conductance histograms \textbf{2-SMe} and \textbf{3-meta}}
\begin{figure}[H]
\centering
\includegraphics[width=\linewidth]{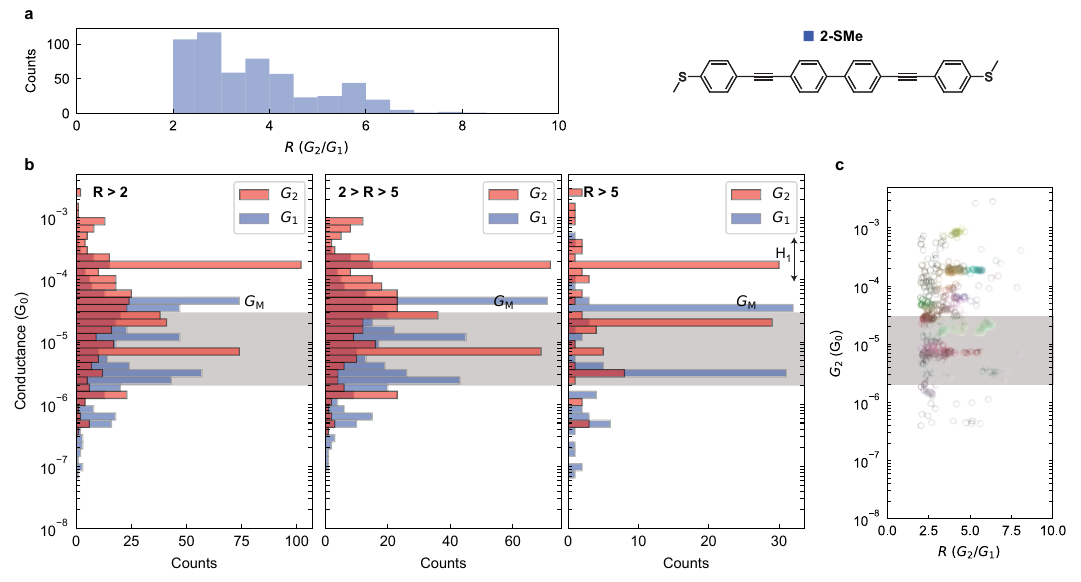}
\caption{Conductance statistics for \textbf{2-SMe}, analogous to the \textbf{1-SAc} analysis in main text Fig. 5.}
\label{conductance_histo_SMe_appendix}
\end{figure}

\begin{figure}[H]
\centering
\includegraphics[width=\linewidth]{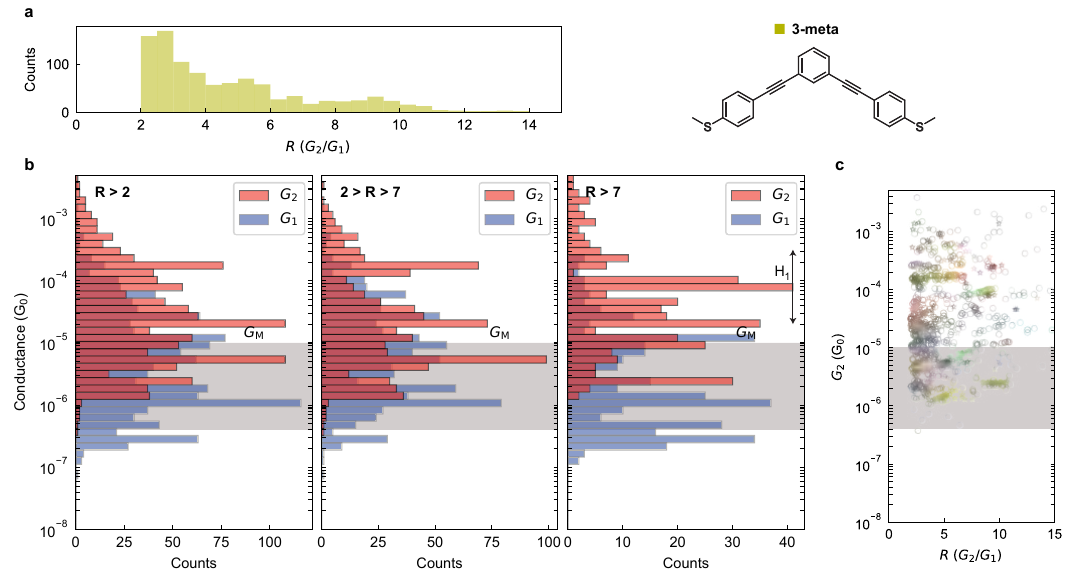}
\caption{Conductance statistics for \textbf{3–meta}, analogous to the \textbf{1-SAc} analysis in main text Fig. 5.}
\label{conductance_histo_meta_appendix}
\end{figure}

The conductance statistics for \textbf{2-SMe} and \textbf{3-meta} are shown in \cref{conductance_histo_SMe_appendix} and \ref{conductance_histo_meta_appendix}, respectively. The same processing workflow as in the main text for \textbf{1-SAc} (Fig. 5) is applied. Based on the ratio distributions in panel~a, we define the high-ratio family using thresholds of $R>5$ for \textbf{2-SMe} and $R>7$ for \textbf{3-meta}. For \textbf{2-SMe}, the conductance histograms in \cref{conductance_histo_SMe_appendix}b exhibit sharper peaks, reflecting the smaller number of memristive events in the dataset. Comparing the peak positions to the room-temperature reference window $G_{\mathrm{M}}$ highlights an additional feature near $10^{-4}\,\mathrm{G}_{0}$, i.e., above the fully stretched conductance. This high-conductance state is compatible with the long-short injection scenario (H$_1$) discussed in the main text. For \textbf{3-meta}, broader conductance distributions are observed, consistent with a larger number of switching events. Both $G_{1}$ and $G_{2}$ show substantial counts within the room-temperature window $G_{\mathrm{M}}$, and, as for \textbf{2-SMe}, an additional contribution above $G_{\mathrm{M}}$ is present, again consistent with $G_\mathrm{M} \leftrightarrow \mathrm{H}_1$ switching.

\section{Threshold Voltage}
\begin{figure}[H]
\centering
\includegraphics[width=\linewidth]{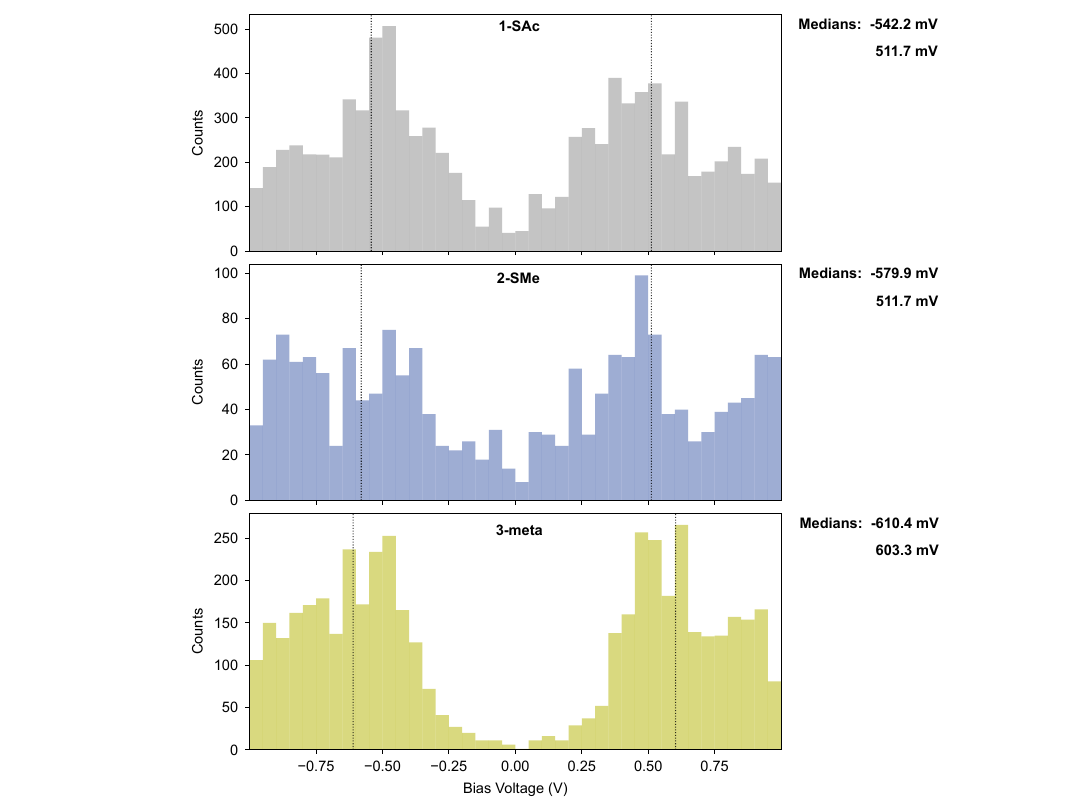}
\caption{Threshold voltage distribution for the three target molecules.}
\label{Threshold_voltage_appendix}
\end{figure}

Figure \ref{Threshold_voltage_appendix} shows the distribution of switching voltage thresholds for the three molecules.
The threshold is defined as the bias closest to zero at which a transition is first observed (i.e., the smallest $|V|$ that triggers switching).
For all molecules, the distributions are approximately symmetric about zero, with medians between $500$ and $700\,\mathrm{mV}$.
\textbf{2–SMe} exhibits two peaks, whereas the other two molecules are predominantly singly peaked with tails that decay at higher bias.
The apparent bimodality for \textbf{2–SMe} likely reflects limited switching statistics, which increases the influence of a few hold positions on the histogram.

\begin{figure}[H]
\centering
\includegraphics[width=\linewidth]{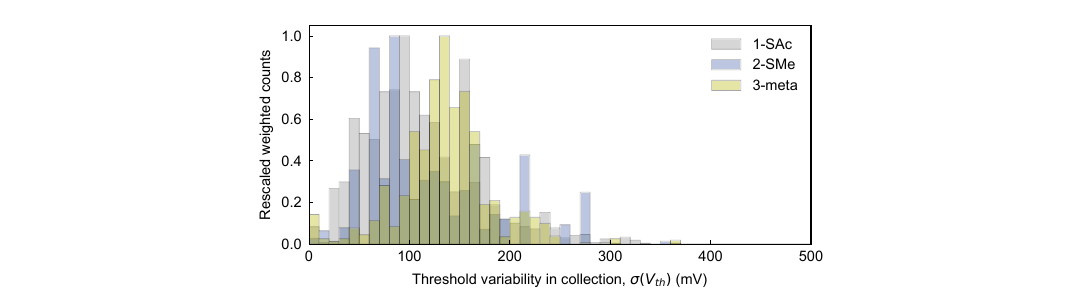}
\caption{ Threshold voltage variability within the same IV collection. The variability is weighted by the number of IVs in each collection, and the histograms are normalized.}
\label{threshold_variability}
\end{figure}

Figure~\ref{threshold_variability} shows the standard deviation of the threshold voltage within a given IV collection, $\sigma(V_{\mathrm{th}})$, for the three molecules. The variability is weighted by the number of IVs in each collection, and the histograms are normalized. All three molecules display a main peak at finite variability. For \textbf{1--SAc} and \textbf{2--SMe}, this peak is centered around $\sigma(V_{\mathrm{th}})\approx 100\,\mathrm{mV}$, whereas for \textbf{3--meta} it is shifted to $\sigma(V_{\mathrm{th}})\approx 150\,\mathrm{mV}$. This provides further support for the more stochastic nature of switching in the meta-connected molecule. In addition to the main peak, the low-$\sigma(V_{\mathrm{th}})$ tail, approaching $\sigma(V_{\mathrm{th}})\approx 0$, corresponds to collections with highly reproducible threshold voltages. In contrast to \textbf{3--meta}, both \textbf{1--SAc} and \textbf{2--SMe} exhibit several low-variability collections, consistent with a more field-driven switching mechanism. The overall behavior of \textbf{1--SAc} and \textbf{2--SMe} is similar, although the distribution for \textbf{1--SAc} is smoother because of the larger number of memristive IVs in the dataset.

\begin{figure}[H]
\centering
\includegraphics[width=0.9\linewidth]{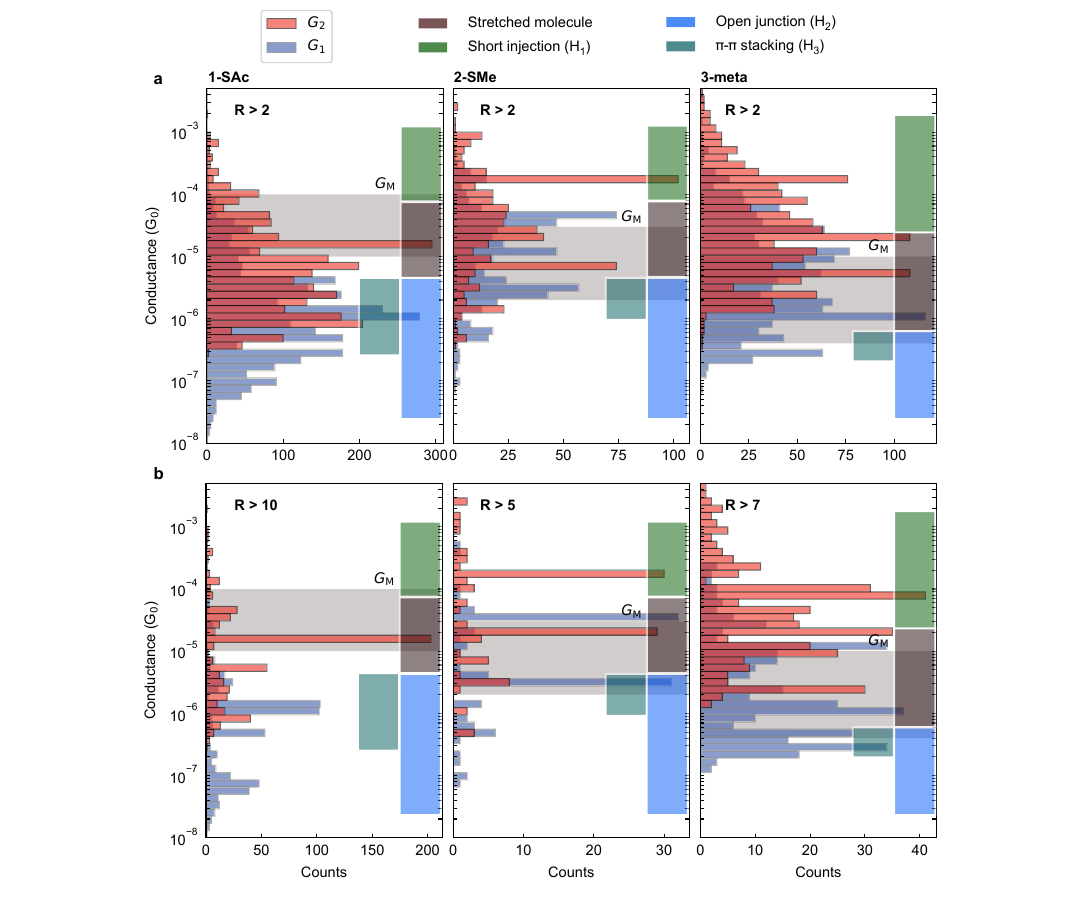}
\caption{Interpreting the origin of switching. (a) Conductance histograms for all memristive switches ($R>2$). (b) Conductance histograms restricted to high-ratio events ($R>10$, $R>5$, and $R>7$, respectively). In all panels, the room-temperature conductance window $\mathrm{G}_{M}$ is depicted as a shaded black rectangle. Colored side bars annotate a proposed microscopic assignment of the conductance states: fully stretched single-molecule junction; molecular junction with injection on the conjugated backbone; $\pi$–$\pi$ stacked multi-molecule junction; and an open junction with one anchor detached from the Au electrode.}
\label{figure5-comparison}
\end{figure}

\section{Switching origin interpretation}
Figure \ref{figure5-comparison} compares the conductance histograms of the three molecules for (i) all memristive IVs ($R>2$) and (ii) the high-ratio subset, defined as $R>10$ for \textbf{1-SAc}, $R>5$ for \textbf{2-SMe}, and $R>7$ for \textbf{3-meta} (see Figs.~5, \ref{conductance_histo_SMe_appendix}, and \ref{conductance_histo_meta_appendix}). Colored bars on the right of each panel indicate the microscopic interpretation assigned to the corresponding conductance states, following the discussion in the main text.

\section{Single-level model} 
\begin{figure}[H]
\centering
\includegraphics[width=\linewidth]{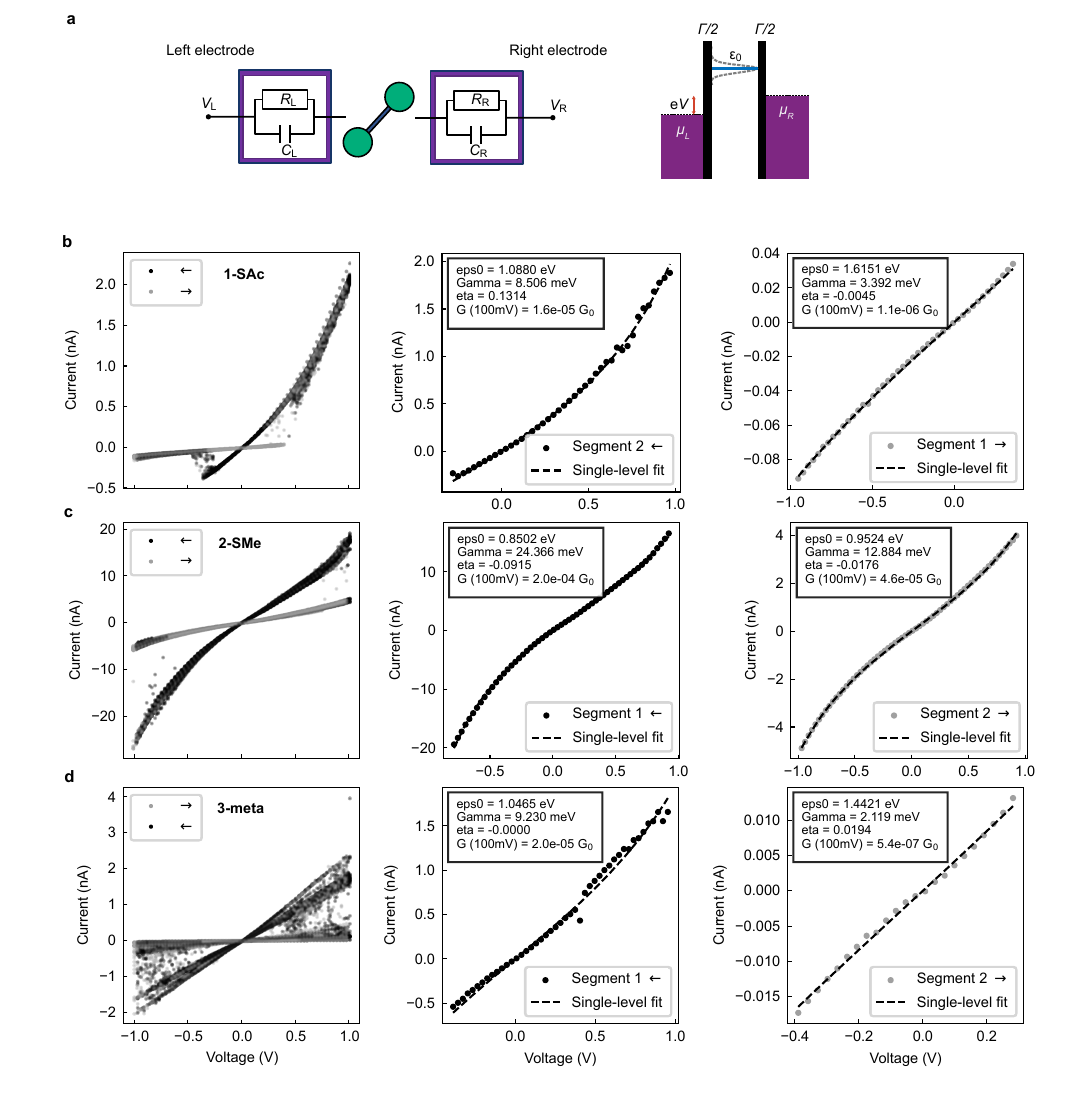}
\caption{ Schematic representation of the model.
(b–d) The left panels show the three collections; the central and right panels show the fits to the unidirectional segments. The free parameters of the model are the zero-bias level position $\varepsilon_0$, the total level broadening $\Gamma$, and the electrostatic asymmetry parameter $\eta$.}
\label{SLM}
\end{figure}

Transport through molecular junctions is often described by the single-level Landauer model. In this framework, transport is dominated by a single molecular orbital whose spectral width is broadened by the coupling to the leads (see \cref{SLM}a). The electrostatic coupling to the left and right electrodes is modeled by an effective capacitive division that is assumed to be bias independent. Importantly, this model does not include any switching mechanism. Therefore, for memristive IVs we treat the two sweep directions separately and fit only the low-bias segments before any switching event occurs. In addition, the single-level model neglects dipolar effects, multi-level alignment, and quantum-interference contributions. As a result, the IV collections in Sec.~I that display more complex behavior (e.g., negative differential conductance or strong asymmetric nonlinearities) cannot be captured within this minimal description. For this reason, we focus on the comparatively simple IV traces shown in Fig.~3, and the corresponding fit parameters are summarized in Fig.~\ref{SLM}. 

\subsection*{Model} 
The current is written as 
\begin{equation} I(V)=-\frac{2e}{h}\int_{-\infty}^{+\infty} T(E,V)\left[f(E,\mu_L)-f(E,\mu_R)\right]\,dE, 
\end{equation} 
where 
\begin{equation} f(E,\mu)=\frac{1}{1+\exp\!\left[(E-\mu)/(k_{\mathrm B}T)\right]} 
\end{equation}
is the Fermi function and the electrode chemical potentials are taken as \begin{equation}
\mu_L=-\frac{eV}{2}, \qquad \mu_R=+\frac{eV}{2}. 
\end{equation} 
The transmission function is 
\begin{equation} 
T(E,V)=\frac{\Gamma_L\Gamma_R}{\left[E-\varepsilon(V)\right]^2+\left(\Gamma/2\right)^2}, 
\end{equation} with total broadening 
\begin{equation} 
\Gamma=\Gamma_L+\Gamma_R. 
\end{equation} Assuming symmetric tunneling coupling to reduce the number of free parameters, we set 
\begin{equation} 
\Gamma_L=\Gamma_R=\frac{\Gamma}{2}, 
\end{equation} so that the fit contains three free parameters: the zero-bias level position $\varepsilon_0$, the total level broadening $\Gamma$, and the electrostatic asymmetry parameter 
\begin{equation}
\eta = \frac{C_L-C_R}{C_L+C_R}. 
\end{equation} 
The parameter $\eta$ describes the shift of the molecular level under bias due to unequal capacitive coupling to the two electrodes. The bias-dependent level position is written as 
\begin{equation} 
\varepsilon(V)=\varepsilon_0-\eta eV. 
\end{equation} 
We fit segments 1 and 2 independently, yielding separate parameter sets for the two conductance states. It is important to note that the sign of $\varepsilon_0$ and $\eta$ cannot be unambiguously assigned because the current satisfies the symmetry
\begin{equation} 
I(V, \Gamma, \varepsilon_0, \eta) = I(V, \Gamma, -\varepsilon_0, -\eta).
\end{equation} 
\subsection*{Interpretation of representative IV collections} 

Figures~\ref{SLM}b-d illustrate three collections that are well captured by the single-level model.
Figure~\ref{SLM}b shows an IV collection whose high-conductance state, with $G(100\,\mathrm{mV})\sim 10^{-5}\,\mathrm{G}_{0}$, is compatible with a fully stretched molecular junction ($G_{\mathrm{M}}$) or a small asymmetric contact rearrangement (L$_1$). The low-conductance state is compatible with an open junction (H$_2$) or a junction bridged by a $\pi$--$\pi$ stacked dimer (H$_3$). By fitting one of the IVs within the single-level model, the parameters cannot prove these interpretations. However, the low-conductance state yields $\eta\sim 0$, which is consistent with a comparatively symmetric configuration (compatible with H$_2$), whereas the high-conductance state yields $\eta\simeq 0.13$, suggesting a more asymmetric junction configuration than an idealized symmetric, fully stretched case. For these reasons, we interpret Fig.~\ref{SLM}b as H$_2\leftrightarrow$L$_1$. 

Figure~\ref{SLM}c, interpreted as long--short injection switching (H$_1\leftrightarrow G_{\mathrm{M}}$), shows $\eta$ values consistent with this picture: the high-conductance (short) configuration exhibits larger asymmetry, while the low-conductance (stretched) configuration has $\eta\sim 0$. 

Finally, in Fig.~\ref{SLM}d both states yield $\eta\sim 0$, and the low-bias conductance levels are consistent with open--closed switching involving a symmetric fully stretched configuration (H$_2\leftrightarrow G_{\mathrm{M}}$). Interestingly, the fitted level misalignment tends to follow $\varepsilon_0\sim 1$~eV for configurations interpreted as closed molecular junctions, and around $\varepsilon_0\sim 1.5$~eV for configurations interpreted as open junctions. A more detailed modeling of the richer and more diverse IV collections could provide additional insights on the switching mechanism, but this lies beyond the scope of the present work.

\begin{comment}
    \section{Before and after a memristive IV}\label{sec:IV_before_after}
Figure~\ref{IV_before_after}a shows the breaking trace corresponding to the \textbf{2--SMe} hold-position discussed in Fig.~2b and \cref{SLM}c. The vertical dotted line marks the electrode displacement at which the memristive IV collection was recorded. Red and blue markers indicate hold positions at smaller and larger electrode displacements, respectively. The corresponding IV characteristics are shown in \cref{IV_before_after}b. For the memristive hold-position, solid and dotted lines denote the two sweep directions, whereas for the neighboring hold-positions we plot only the sweep toward positive bias for clarity.

Based on the conductance analysis in the main text and the single-level model comparison, the two conductance states are interpreted as long--short injection switching (H$_1\leftrightarrow G_{\mathrm{M}}$). Figure~\ref{IV_before_after}b supports this picture: the high-conductance state appears at smaller electrode displacement (shorter junction), while the low-conductance state remains near $10^{-5}\,\mathrm{G}_{0}$ over a broad displacement range, until the conductance drops to the noise, consistent with a fully stretched single-molecule junction.

\begin{figure}[H]
\centering
\includegraphics[width=\linewidth]{figures_SI/appendix/IV_before_after.pdf}
\caption{ Single level model applied to the memristive IV collections shown in Fig. 2.}
\label{IV_before_after}
\end{figure}

\end{comment}

\section{Memristive IVs}\label{sec:memristive_IVs}
In the following, raw plots of memristive IV collections illustrate the diversity of IV shapes, asymmetries, nonlinearities, and stability, providing complementary information to the zero-bias conductance statistics discussed above. Each row corresponds to one IV collection. Panel~a shows the superposed IVs using the color convention of Fig. 3 in the main text: red for sweeps to the right and gray for sweeps to the left. A text box in the upper-left corner reports the total number of IVs acquired at that hold position, the number classified as memristive according to the criteria in Sec.~\ref{sec:meristive_iv_classification}, and the collection-averaged zero-bias ratio $R=G_{2}/G_{1}$. 

Panel~b shows the conductance--voltage traces $G(V)=I/V$ for all IVs in the collection, whereas panel~c shows only the subset classified as memristive. Panel~d shows the breaking and making traces and marks the displacement at which the IV collection was recorded. Panel~e shows conductance histograms for all memristive IVs ($R> 2$); black dotted lines indicate the collection-averaged conductance levels $G_{1}$ and $G_{2}$.

\subsection*{1-SAc}
\begin{figure}[H]
\centering
\includegraphics[width=\linewidth]{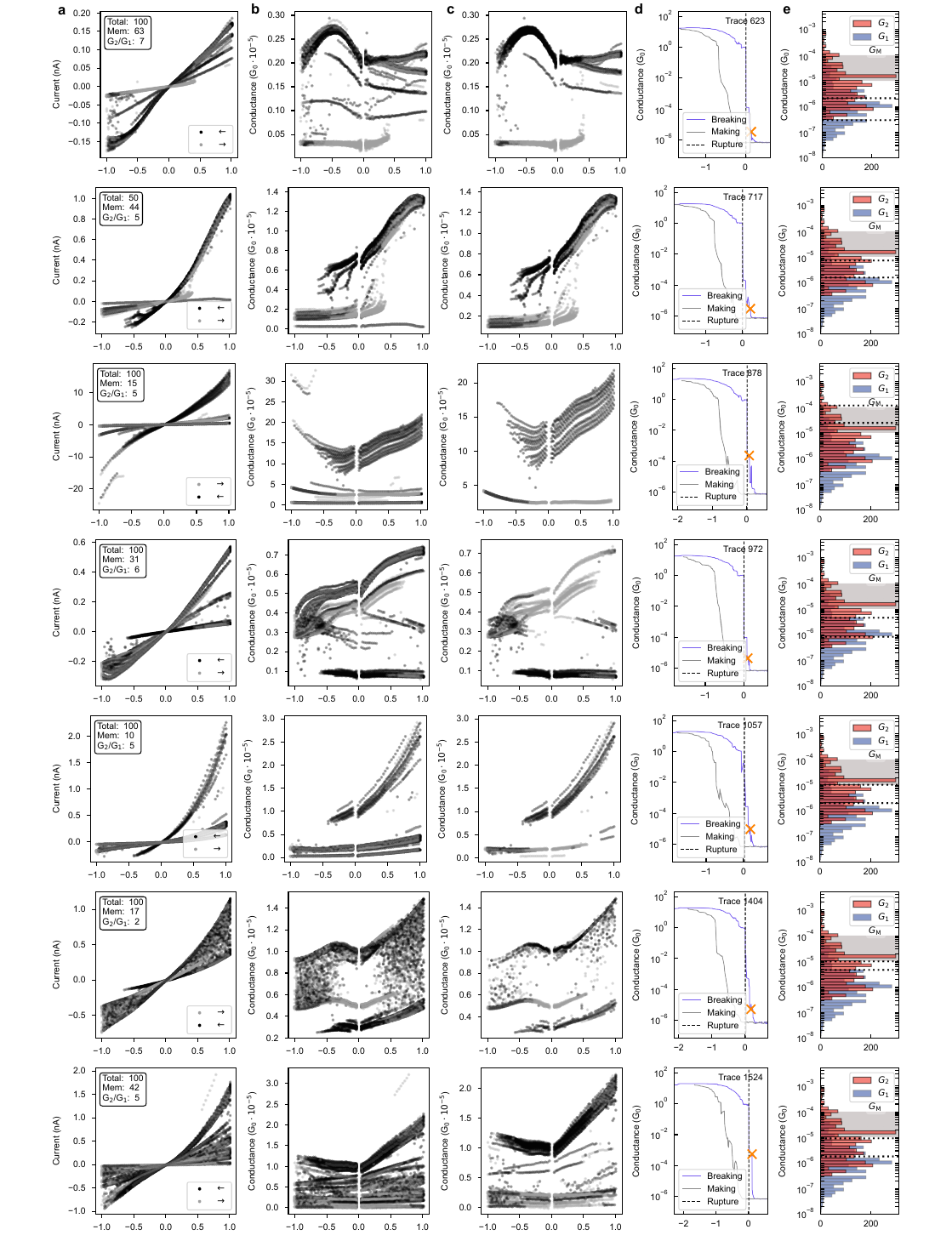}
\label{appendix-1SAc-1}
\end{figure}

\begin{figure}[H]
\centering
\includegraphics[width=\linewidth]{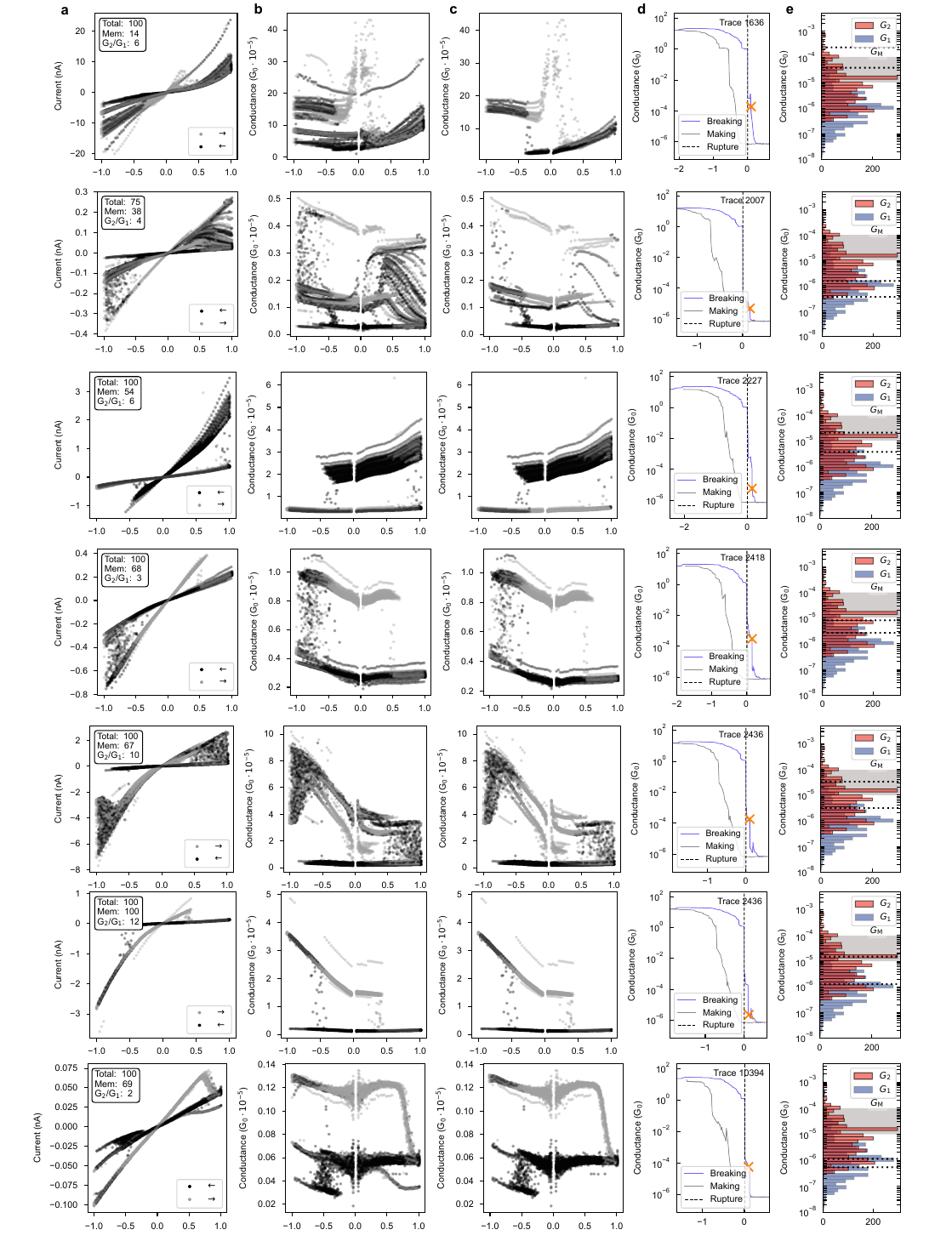}
\label{appendix-1SAc-2}
\end{figure}

\begin{figure}[H]
\centering
\includegraphics[width=\linewidth]{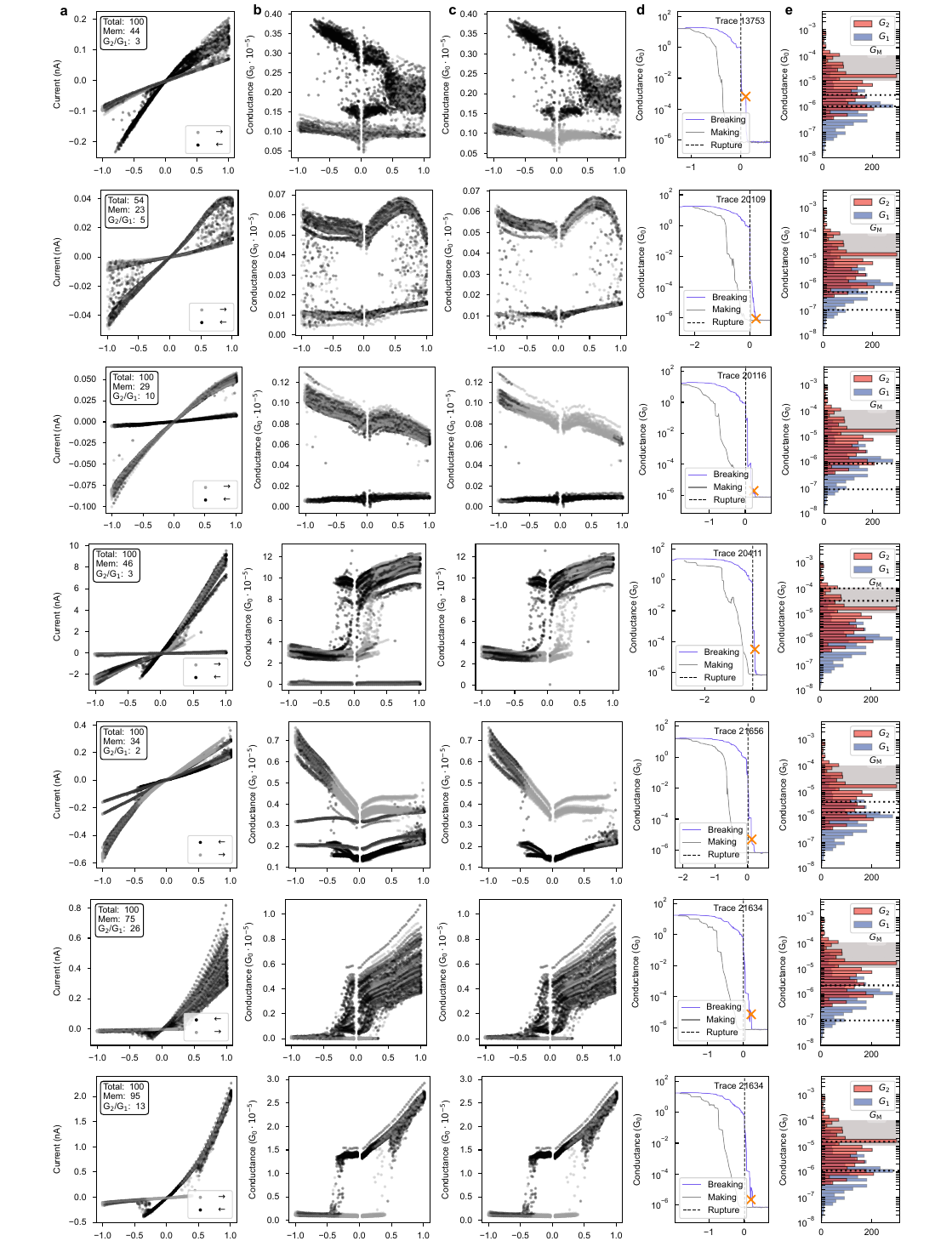}
\label{appendix-1SAc-3}
\end{figure}

\subsection*{2-SMe}
\begin{figure}[H]
\centering
\includegraphics[width=\linewidth]{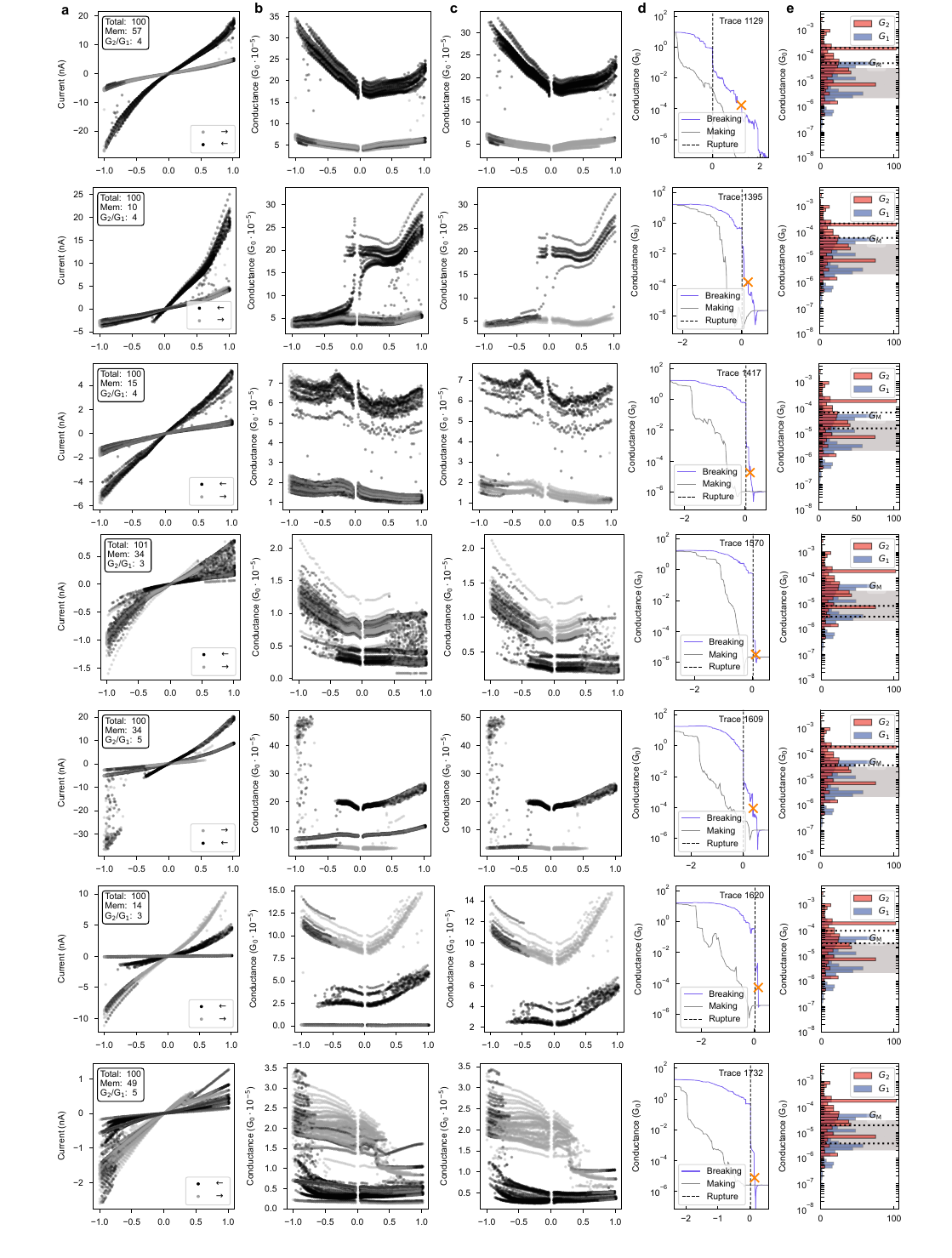}
\label{appendix-2SMe-1}
\end{figure}

\subsection*{3-meta}
\begin{figure}[H]
\centering
\includegraphics[width=\linewidth]{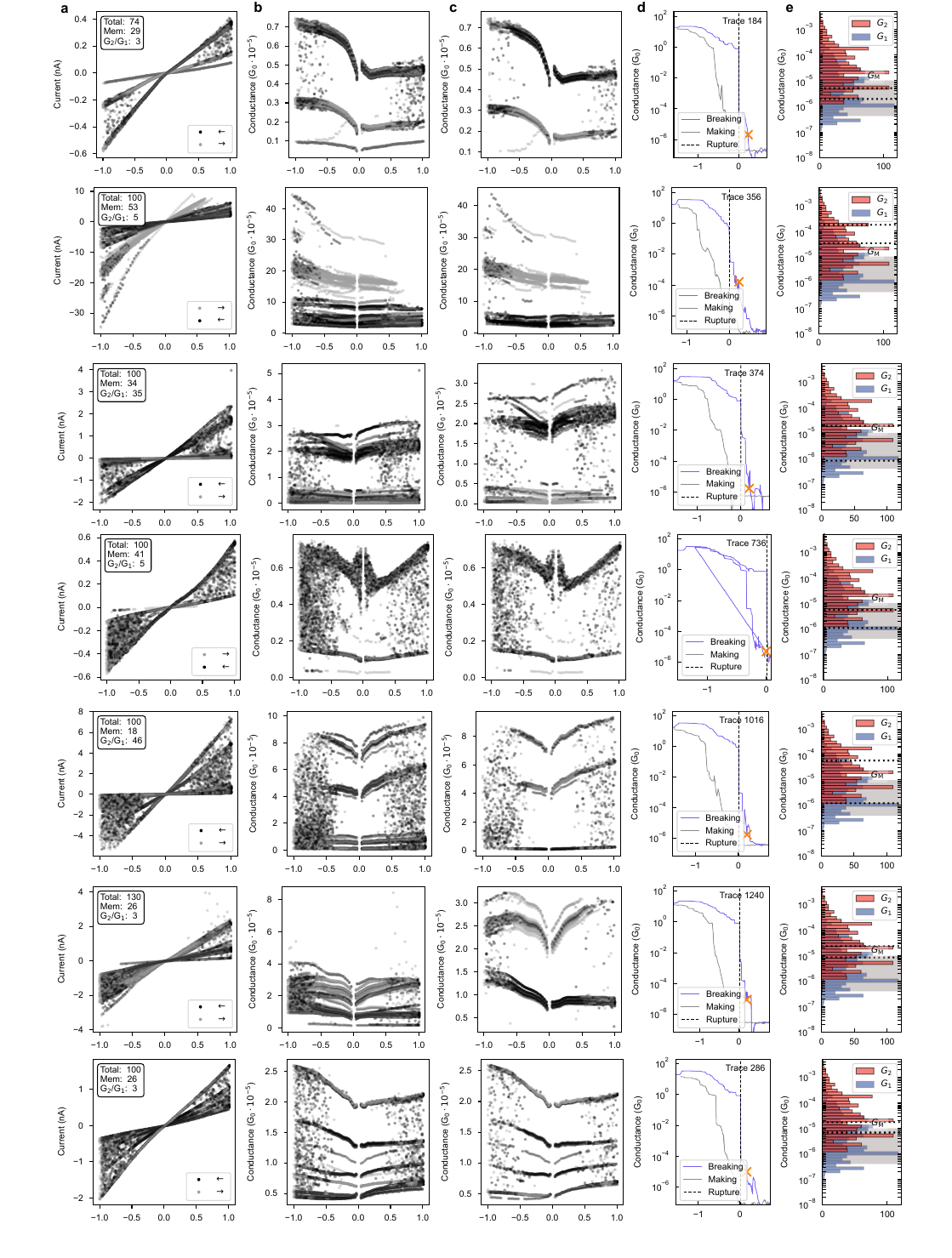}
\label{appendix-3meta-1}
\end{figure}
\section{Materials and General Methods}

\noindent \textbf{Reagents} (Acros, Aldrich, AmBeed, Chemat, and TCI) were purchased as reagent grade and used without further purification.

\vspace{5pt}

\noindent \textbf{Solvents} for extraction or column chromatography were used analytical grade.

\vspace{5pt}

\noindent \textbf{Dry solvents} (THF, \ce{CH2Cl2}, diethyl ether, and toluene) for reactions were purified by a solvent drying system from MBraun under nitrogen atmosphere (\ce{H2O} content $< 10$ ppm as determined by Karl-Fischer titration). All other solvents were purchased in p.a.\ quality.

\vspace{5pt}

\noindent \textbf{Reactions} in the absence of air and moisture were performed in oven-dried glassware under Ar atmosphere.

\vspace{5pt}

\noindent \textbf{Flash column chromatography (FC)} was performed using Biotage\textsuperscript{\tiny\textregistered} Selekt apparatus at \SI{25}{\celsius} with a head pressure of 0.0--30 bar and Flow Rate (50--250 mL/min). \ce{SiO2} (60 \AA, 230--400 mesh, particle size 0.040--0.063 mm, Fluka). The used solvent compositions are reported in synthetic procedures.

\vspace{5pt}

\noindent \textbf{Analytical thin layer chromatography (TLC)} was performed on aluminium sheets coated with silica gel 60 F254 (Merck, Macherey-Nagel). Visualization was achieved using UV light (254 or 365 nm).

\vspace{5pt}

\noindent \textbf{Evaporation in vacuo} was performed at \SIrange{25}{60}{\celsius} and 800--10 mbar.

\vspace{5pt}

\noindent \textbf{Reported yields} refer to spectroscopically and chromatographically pure compounds that were dried under high vacuum (0.5--0.1 mbar) before analytical characterization.

\vspace{5pt}

\noindent \textbf{\textsuperscript{1}H and \textsuperscript{13}C nuclear magnetic resonance (NMR)} spectra were recorded on Bruker 400 (Avance III HD), Bruker DRX 500, Varian-Agilent 500 and Varian-Agilent 600 spectrometers at 400 MHz, 500 MHz or 600 MHz (\textsuperscript{1}H) and 75 MHz, 126 MHz or 150 MHz (\textsuperscript{13}C), respectively. Temperatures of measurements are indicated in the procedures and on the spectra. Chemical shifts $\delta$ are reported in ppm downfield from tetramethylsilane using the residual deuterated solvent signals as an internal reference (CDCl\textsubscript{3}: $\delta\mathrm{H}$ = 7.26 ppm, $\delta\mathrm{C}$ = 77.0 ppm, CD\textsubscript{2}Cl\textsubscript{2}: $\delta\mathrm{H}$ = 5.33 ppm, $\delta\mathrm{C}$ = 53.4 ppm). For \textsuperscript{1}H NMR, coupling constants $J$ are given in Hz and the resonance multiplicity is described as s (singlet), d (doublet), t (triplet), q (quartet), m (multiplet).

\section{Synthetic Protocols}

\subsection*{S-(4-iodophenyl) ethanethioate (1a)}
\begin{figure}[htbp]
  \centering
  \includegraphics[width=0.5\linewidth]{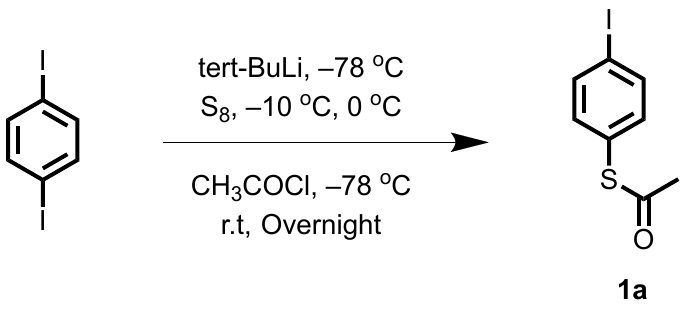}
  \label{fig:1a}
\end{figure}
\noindent Compound \textbf{1a} was synthesized following a modified literature procedure \cite{Hortholary2003AnMonolayers}. tert-Butyllithium (t-BuLi; 16.8 mL, 28.5 mmol, 1.7 M in pentane) was added dropwise to a solution of 1,4-diiodobenzene (5.00 g, 15.2 mmol) in dry diethyl ether (167 mL) at \SI{-78}{\celsius} under an argon atmosphere. The reaction mixture was stirred vigorously at this temperature for 10 min and then allowed to warm gradually. When the internal temperature reached \SI{-10}{\celsius}, elemental sulfur (\ce{S8}, 505 mg, 1.97 mmol) was added in one portion. The reaction mixture was maintained at \SI{0}{\celsius} for 30 min and subsequently cooled again to \SI{-78}{\celsius}, followed by the addition of acetyl chloride (1.5 g, 1.36 mL, 19.1 mmol) in one portion. The reaction mixture was allowed to warm to room temperature and stirred overnight. After aqueous workup, the mixture was extracted with \ce{CH2Cl2}. The combined organic layers were washed with a dilute aqueous \ce{NaHCO3} solution, dried over anhydrous \ce{MgSO4}, filtered, and concentrated under reduced pressure. The crude product was purified by flash column chromatography (\ce{SiO2}, hexanes:ethyl acetate 95:5), followed by recrystallization from hexanes to afford \textbf{1a} (2.6 g, 62\%) as a white solid. \textsuperscript{1}H NMR data are consistent with previously reported values \cite{Hortholary2003AnMonolayers}.

\textbf{\textsuperscript{1}H NMR:} (400 MHz, CDCl\textsubscript{3}) $\delta$ 7.74 (d, $J$ = 8.4 Hz, 2H), 7.13 (d, $J$ = 8.4 Hz, 2H), 2.42 ppm (s, 3H).

\subsection*{4,4'-bis((trimethylsilyl)ethynyl)-1,1'-biphenyl (2a)}
\begin{figure}[htbp]
  \centering
  \includegraphics[width=\linewidth]{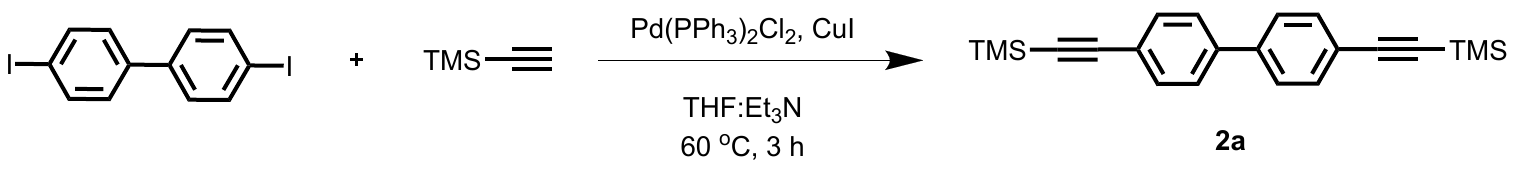}
  \label{fig:2a}
\end{figure}
\noindent Compound \textbf{2a} was synthesized following a modified literature procedure \cite{Olavarria-Contreras2016CAuGroups}.
Ethynyltrimethylsilane (0.80 mL, 5.42 mmol) was added to a degassed solution of 4,4'-diiodo-1,1'-biphenyl (1.00 g, 2.5 mmol) in a mixture of dry THF and triethylamine (90 mL, 60:30 v/v) under an argon atmosphere. The reaction mixture was flushed with argon for an additional 5 min, after which \ce{Pd(PPh3)2Cl2} (86 mg, 0.12 mmol) and \ce{CuI} (23 mg, 0.12 mmol) were added. The reaction mixture was stirred at \SI{60}{\celsius} under argon for 3 h. After cooling to room temperature, the reaction mixture was poured into water and extracted with \ce{CH2Cl2}. The combined organic layers were concentrated under reduced pressure, and the crude residue was purified by flash column chromatography (\ce{SiO2}, hexanes:ethyl acetate 98:2) to afford \textbf{2a} (0.81 g, 95\%) as a light-yellow crystalline solid. Spectroscopic data are consistent with previously reported values \cite{Olavarria-Contreras2016CAuGroups}.

\textbf{\textsuperscript{1}H NMR:} (400 MHz, CDCl\textsubscript{3}) $\delta$ 7.52 (s, 8H), 0.26 ppm (s, 18H).

\subsection*{4,4'-diethynyl-1,1'-biphenyl (2b)}
\begin{figure}[htbp]
  \centering
  \includegraphics[width=\linewidth]{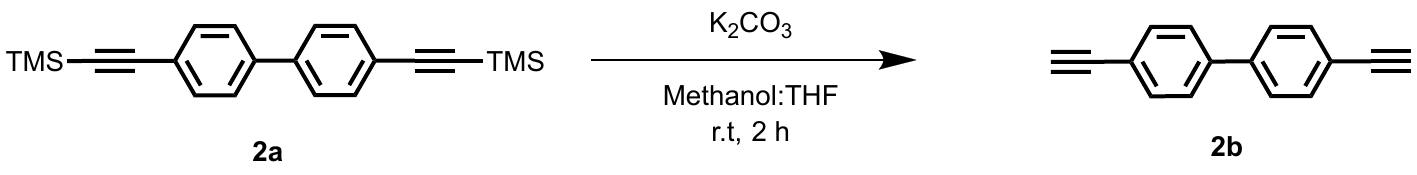}
  \label{fig:2b}
\end{figure}
\noindent Compound \textbf{2b} was synthesized following a modified literature procedure \cite{Olavarria-Contreras2016CAuGroups}. 4,4'-Bis((trimethylsilyl)ethynyl)-1,1'-biphenyl (1.00 g, 2.9 mmol) and \ce{K2CO3} (2.00 g, 14.4 mmol) were added to a mixture of dry THF and methanol (168 mL, 134:34 v/v) at \SI{25}{\celsius}. The reaction mixture was stirred for 2 h, after which \ce{CH2Cl2} was added. The resulting mixture was passed through a short pad of silica gel and washed with \ce{CH2Cl2}. The solvents were removed under reduced pressure to afford the crude product. The residue was purified by flash column chromatography (\ce{SiO2}, hexanes:ethyl acetate 9:1) to afford \textbf{2b} (0.38 g, 65\%) as a white powder. \textsuperscript{1}H NMR data are consistent with previously reported values \cite{Olavarria-Contreras2016CAuGroups}.

\textbf{\textsuperscript{1}H NMR:} (400 MHz, CDCl\textsubscript{3}) $\delta$ 7.58--7.53 (m, 8H), 3.14 ppm (s, 2H).

\subsection*{S,S'-(([1,1'-biphenyl]-4,4'-diylbis(ethyne-2,1-diyl))bis(4,1-phenylene)) diethanethioate (1-SAc)}
\begin{figure}[htbp]
  \centering
  \includegraphics[width=\linewidth]{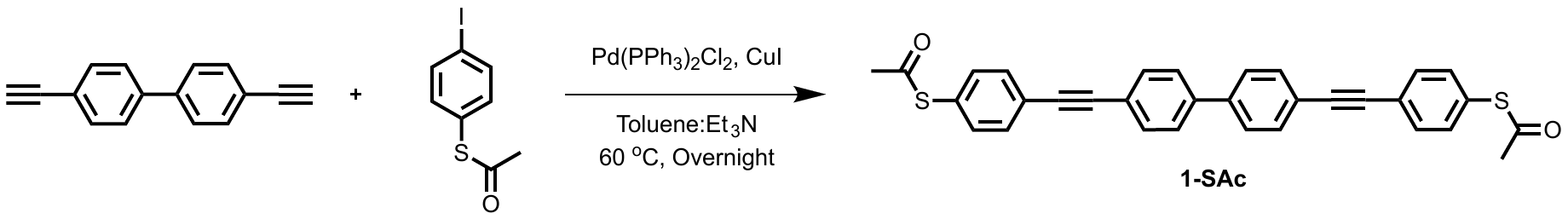}
  \label{fig:1-SAc}
\end{figure}
\noindent \textbf{1-SAc} was synthesized following a modified literature procedure \cite{Soni2020UnderstandingJunctions}. \textbf{2b} (120 mg, 0.6 mmol) and \textbf{1a} (380 mg, 1.36 mmol) were dissolved in a mixture of freshly distilled triethylamine (8 mL) and anhydrous toluene (6 mL). The solution was degassed by purging with argon, after which \ce{Pd(PPh3)2Cl2} (8.3 mg, 0.012 mmol) and \ce{CuI} (5.6 mg, 0.03 mmol) were added. The reaction mixture was further purged with argon for 5 min, then sealed and stirred at \SI{60}{\celsius} overnight under an argon atmosphere. Due to the poor solubility of the product in toluene, the reaction mixture was filtered directly. The filtrate was passed through a short pad of silica gel and eluted with \ce{CH2Cl2}. The solvents were removed under reduced pressure, and the crude product was recrystallized from \ce{CH2Cl2}/hexanes to afford \textbf{1-SAc} (180 mg, 60\%) as an off-white powder. \textsuperscript{1}H and \textsuperscript{13}C NMR spectra are consistent with previously reported values \cite{Soni2020UnderstandingJunctions}.

\textbf{\textsuperscript{1}H NMR:} (500 MHz, CDCl\textsubscript{3}) $\delta$ 7.61 (s, 8H), 7.57 (d, $J$ = 8.4 Hz, 4H), 7.41 (d, $J$ = 8.4 Hz, 4H), 2.44 ppm (s, 6H).

\textbf{\textsuperscript{13}C NMR:} (126 MHz, CDCl\textsubscript{3}) $\delta$ 193.42, 140.25, 134.23, 132.20, 132.17, 128.13, 126.93, 124.47, 122.26, 90.86, 89.64, 30.28 ppm.

\subsection*{4,4'-bis((4-(methylthio)phenyl)ethynyl)-1,1'-biphenyl (2-SMe)}
\begin{figure}[htbp]
  \centering
  \includegraphics[width=\linewidth]{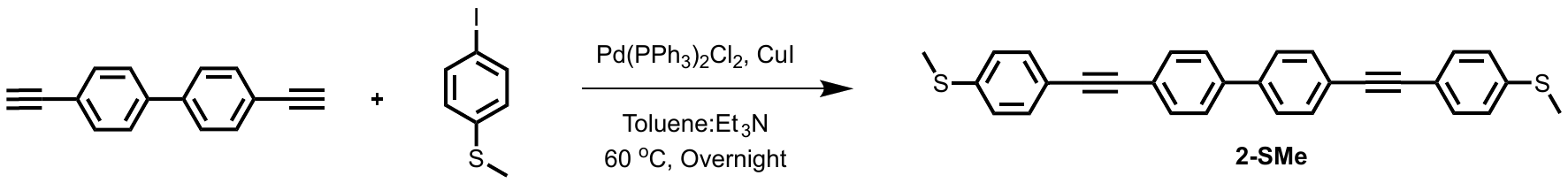}
  \label{fig:2-SMe}
\end{figure}
\noindent \textbf{2-SMe} was synthesized following a modified literature procedure \cite{Gantenbein2019ExploringDerivatives}. \textbf{2b} (100 mg, 0.5 mmol) and (4-iodophenyl)(methyl)sulfane (284 mg, 1.1 mmol) were dissolved in a mixture of dry toluene (7.7 mL) and triethylamine (1.6 mL) and the resulting solution was degassed by purging with argon for 15 min. \ce{CuI} (11.3 mg, 0.06 mmol) and \ce{Pd(PPh3)2Cl2} (35 mg, 0.05 mmol) were then added, and the reaction mixture was further purged with argon for 5 min. The reaction vessel was sealed and stirred at \SI{60}{\celsius} under an argon atmosphere overnight. After cooling to room temperature, the reaction mixture was filtered and washed with methanol. The crude product was purified by flash column chromatography (\ce{SiO2}, hexanes:\ce{CH2Cl2} 3:1) to afford \textbf{2-SMe} (140 mg, 63\%) as a yellow solid, which was further purified by recrystallization from toluene/hexanes. \textsuperscript{1}H and \textsuperscript{13}C NMR spectra are consistent with previously reported values. The \textsuperscript{13}C NMR spectrum was recorded in CDCl\textsubscript{3}, where limited solubility resulted in low signal intensity. Despite this, the number of observed resonances was consistent with the proposed structure. DMSO-d\textsubscript{6} was also evaluated as an alternative solvent but did not result in a significant improvement in spectral quality. Although a DMSO-d\textsubscript{6}/\ce{CS2} 1:1 (v/v) solvent mixture has been reported for NMR characterization in ref.\ \cite{Gantenbein2019ExploringDerivatives}, it was not employed in the present study due to handling considerations, as structural assignment could be achieved without its use.

\textbf{\textsuperscript{1}H NMR:} (500 MHz, CDCl\textsubscript{3}) $\delta$ 7.60 (s, 8H), 7.46 (d, $J$ = 8.1 Hz, 4H), 7.22 (d, $J$ = 8.1 Hz, 4H), 2.51 ppm (s, 6H).

\textbf{\textsuperscript{13}C NMR:} (151 MHz, CDCl\textsubscript{3}) $\delta$ 146.40, 132.01, 131.86, 130.17, 126.84, 125.87, 124.80, 118.73, 96.10, 90.20, 15.37 ppm.

\subsection*{1,3-bis((trimethylsilyl)ethynyl)benzene (3a)}
\begin{figure}[htbp]
  \centering
  \includegraphics[width=0.7\linewidth]{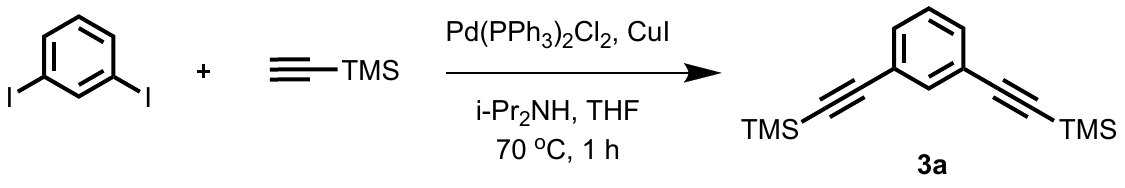}
  \label{fig:3a}
\end{figure}
\noindent Compound \textbf{3a} was synthesized following a modified literature procedure \cite{Lee2010IntramolecularBinding}. 1,3-Diiodobenzene (500 mg, 1.52 mmol) and i-Pr\textsubscript{2}NH (1.23 g, 1.71 mL, 12.2 mmol) were dissolved in anhydrous THF (9 mL), and the resulting solution was degassed by purging with argon for 10 min. \ce{Pd(PPh3)2Cl2} (32 mg, 0.045 mmol) and \ce{CuI} (35 mg, 0.18 mmol) were then added, followed by ethynyltrimethylsilane (328 mg, 3.34 mmol). The reaction mixture was stirred at \SI{70}{\celsius} for 1 h. After cooling to room temperature, the reaction mixture was filtered through a pad of Celite and concentrated under reduced pressure. The crude product was purified by flash column chromatography (\ce{SiO2}, hexanes:ethyl acetate 10:1) to afford \textbf{3a} (403 mg, 98\%) as a white solid. \textsuperscript{1}H spectra are consistent with previously reported values \cite{Lee2010IntramolecularBinding}.

\textbf{\textsuperscript{1}H NMR:} (400 MHz, CDCl\textsubscript{3}) $\delta$ 7.58 (t, $J$ = 1.4 Hz, 1H), 7.38 (dd, $J$ = 7.8, 1.4 Hz, 2H), 7.24 (t, $J$ = 1.4 Hz, 1H), 0.24 ppm (s, 18H).

\subsection*{1,3-diethynylbenzene (3b)}
\begin{figure}[htbp]
  \centering
  \includegraphics[width=0.5\linewidth]{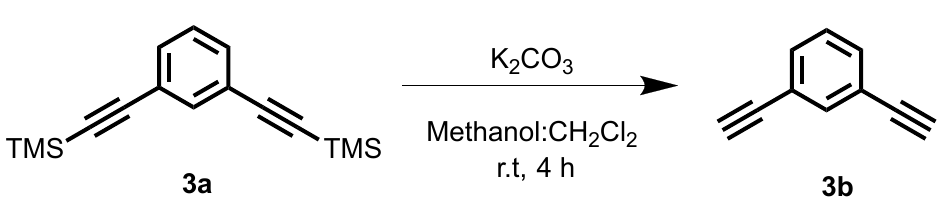}
  \label{fig:3b}
\end{figure}
\noindent Compound \textbf{3b} was synthesized following a modified literature procedure \cite{Ezenwafor2025SupramolecularPrinciples}. Potassium carbonate (2.04 g, 15.0 mmol) was added to a solution of \textbf{3a} (0.40 g, 1.5 mmol) in a mixture of \ce{CH2Cl2} and methanol (39 mL, 2:1 v/v) at \SI{25}{\celsius}. The reaction mixture was stirred at \SI{25}{\celsius} for 4 h, after which water (50 mL) was added. The organic layer was extracted with \ce{CH2Cl2} and the combined organic extracts were dried over anhydrous \ce{MgSO4}. The solvent was removed under reduced pressure to afford \textbf{3b} (145 mg, 78\%) as a brown liquid. Spectroscopic data are consistent with previously reported values \cite{Ezenwafor2025SupramolecularPrinciples}.

\textbf{\textsuperscript{1}H NMR:} (400 MHz, CDCl\textsubscript{3}) $\delta$ 7.61 (t, $J$ = 1.6 Hz, 1H), 7.46 (dd, $J$ = 7.7, 1.6 Hz, 2H), 7.29 (t, $J$ = 7.7 Hz, 1H), 3.08 ppm (s, 2H).

\subsection*{1,3-bis((4-(methylthio)phenyl)ethynyl)benzene (3-meta)}
\begin{figure}[htbp]
  \centering
  \includegraphics[width=1\linewidth]{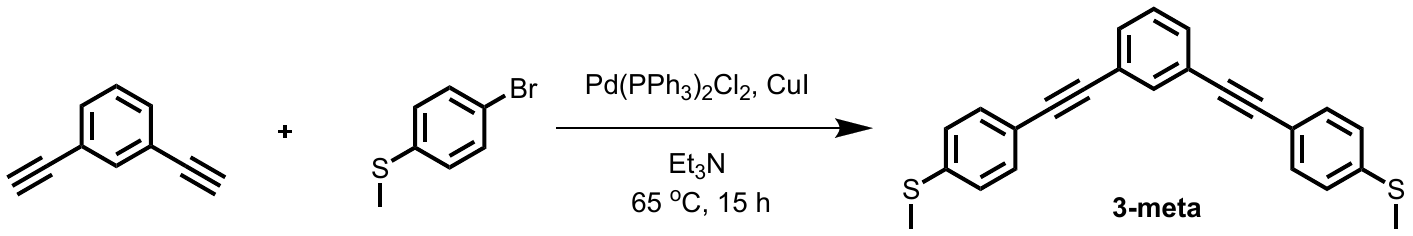}
  \label{fig:3-meta}
\end{figure}
In contrast to the reported procedure \cite{Yan2010Molecularly-mediatedStructures}, which employed a stepwise Sonogashira coupling of (4-ethynylphenyl)(methyl)sulfane with 1,3-diiodobenzene under slow addition conditions, the target compound was synthesized in the present work via an alternative synthetic strategy. Specifically, the corresponding diethynylbenzene derivative \textbf{3b} was prepared first and subsequently coupled with the aryl bromide under standard Sonogashira conditions. This approach avoids prolonged reagent addition and allows the final coupling to proceed in a single step, while affording the desired product in comparable yield.

\textbf{3b} (360 mg, 2.8 mmol) and (4-bromophenyl)(methyl)sulfane (1.45 g, 7.1 mmol) were dissolved in freshly distilled triethylamine (10 mL). The solution was degassed by purging with argon, after which \ce{Pd(PPh3)2Cl2} (160 mg, 0.23 mmol) and \ce{CuI} (54 mg, 0.28 mmol) were added. The reaction mixture was further purged with argon for 5 min and then stirred at \SI{65}{\celsius} for 15 h under an argon atmosphere. After cooling to room temperature, the reaction mixture was washed with sat.\ aqueous \ce{NH4Cl} and extracted with ethyl acetate. The combined organic layers were washed with brine, dried over anhydrous \ce{MgSO4}, filtered, and concentrated under reduced pressure. The residue was purified by flash column chromatography (\ce{SiO2}, hexanes:ethyl acetate 95:5) followed by recrystallization from methanol/\ce{CH2Cl2} to afford \textbf{3-meta} (800 mg, 76\%) as a white crystalline solid. \textsuperscript{1}H and \textsuperscript{13}C NMR spectra are consistent with previously reported values \cite{Yan2010Molecularly-mediatedStructures}.

\textbf{\textsuperscript{1}H NMR:} (500 MHz, CD\textsubscript{2}Cl\textsubscript{2}) $\delta$ 7.69 (s, 1H), 7.50 (dd, $J$ = 7.7, 1.7 Hz, 2H), 7.47 (d, $J$ = 8.4 Hz, 4H), 7.37 (t, $J$ = 7.7 Hz, 1H), 7.24 (d, $J$ = 8.4 Hz, 4H), 2.52 ppm (s, 6H).

\textbf{\textsuperscript{13}C NMR:} (126 MHz, CD\textsubscript{2}Cl\textsubscript{2}) $\delta$ 140.00, 134.17, 131.82, 131.08, 128.59, 125.68, 123.67, 118.98, 89.69, 88.38, 15.02 ppm.

\section{NMR Spectra}

% Figures placeholders (PDF pages 7--12)
\begin{figure}[htbp]
  \centering
  \includegraphics[width=0.8\linewidth]{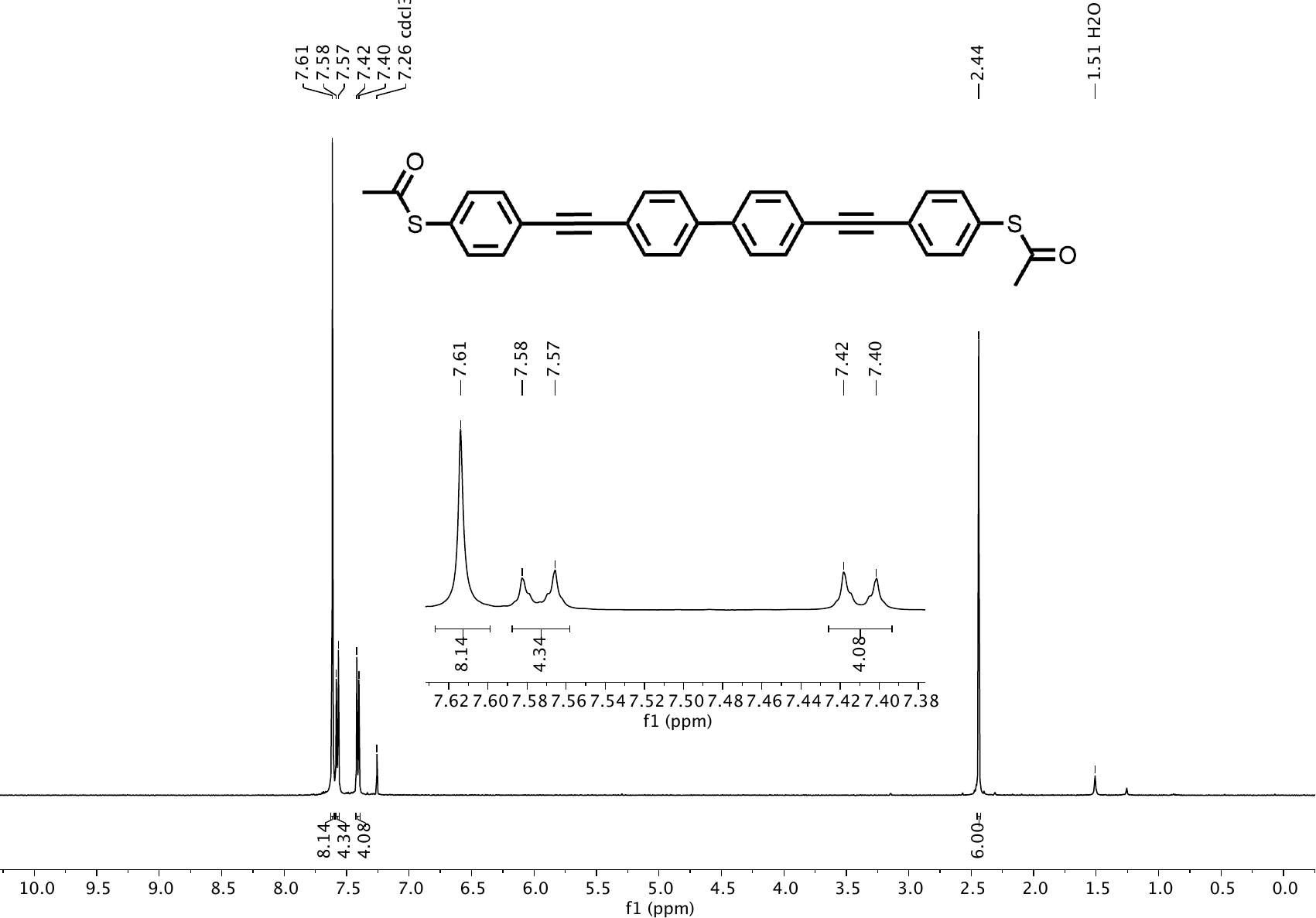}
  \caption{\textsuperscript{1}H NMR (500 MHz) spectra of \textbf{1-SAc} in CDCl\textsubscript{3}.}
  \label{fig:nmr1}
\end{figure}

\begin{figure}[htbp]
  \centering
    \includegraphics[width=0.8\linewidth]{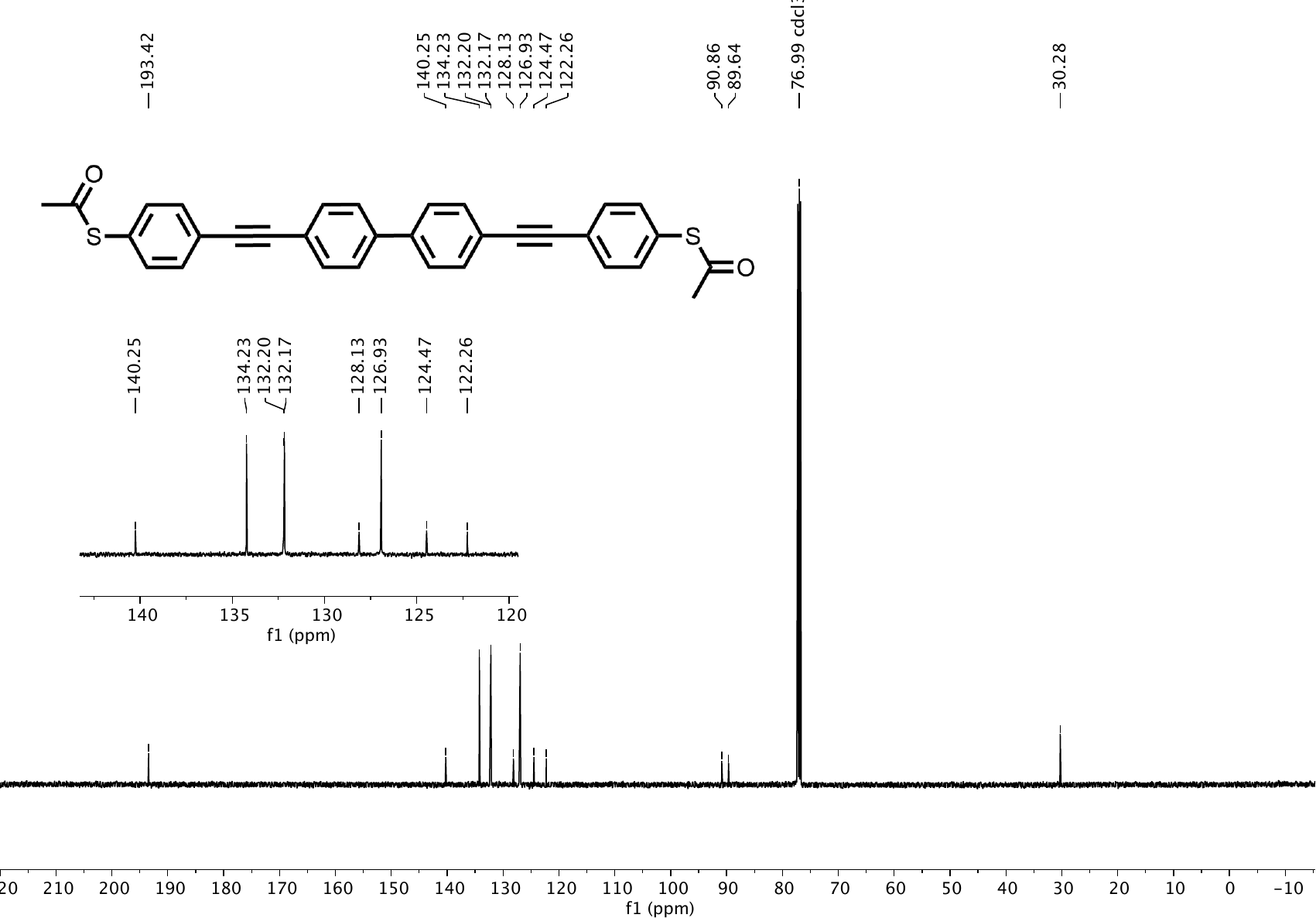}
  \caption{\textsuperscript{13}C NMR (500 MHz) spectra of \textbf{1-SAc} in CDCl\textsubscript{3}.}
  \label{fig:nmr2}
\end{figure}

\begin{figure}[htbp]
  \centering
    \includegraphics[width=0.8\linewidth]{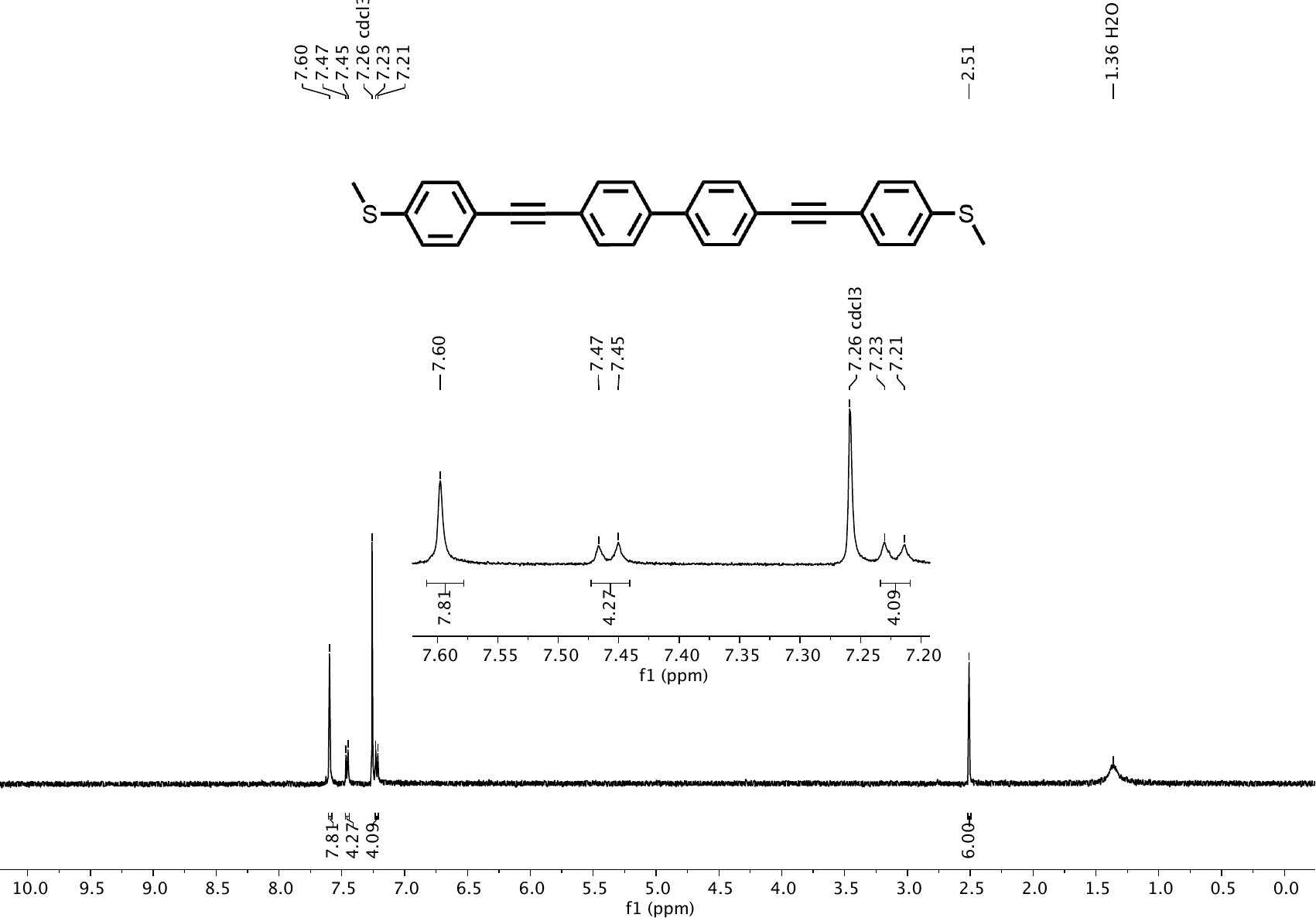}
  \caption{\textsuperscript{1}H NMR (500 MHz) spectra of \textbf{2-SMe} in CDCl\textsubscript{3}.}
  \label{fig:nmr3}
\end{figure}

\begin{figure}[htbp]
  \centering
  \includegraphics[width=0.8\linewidth]{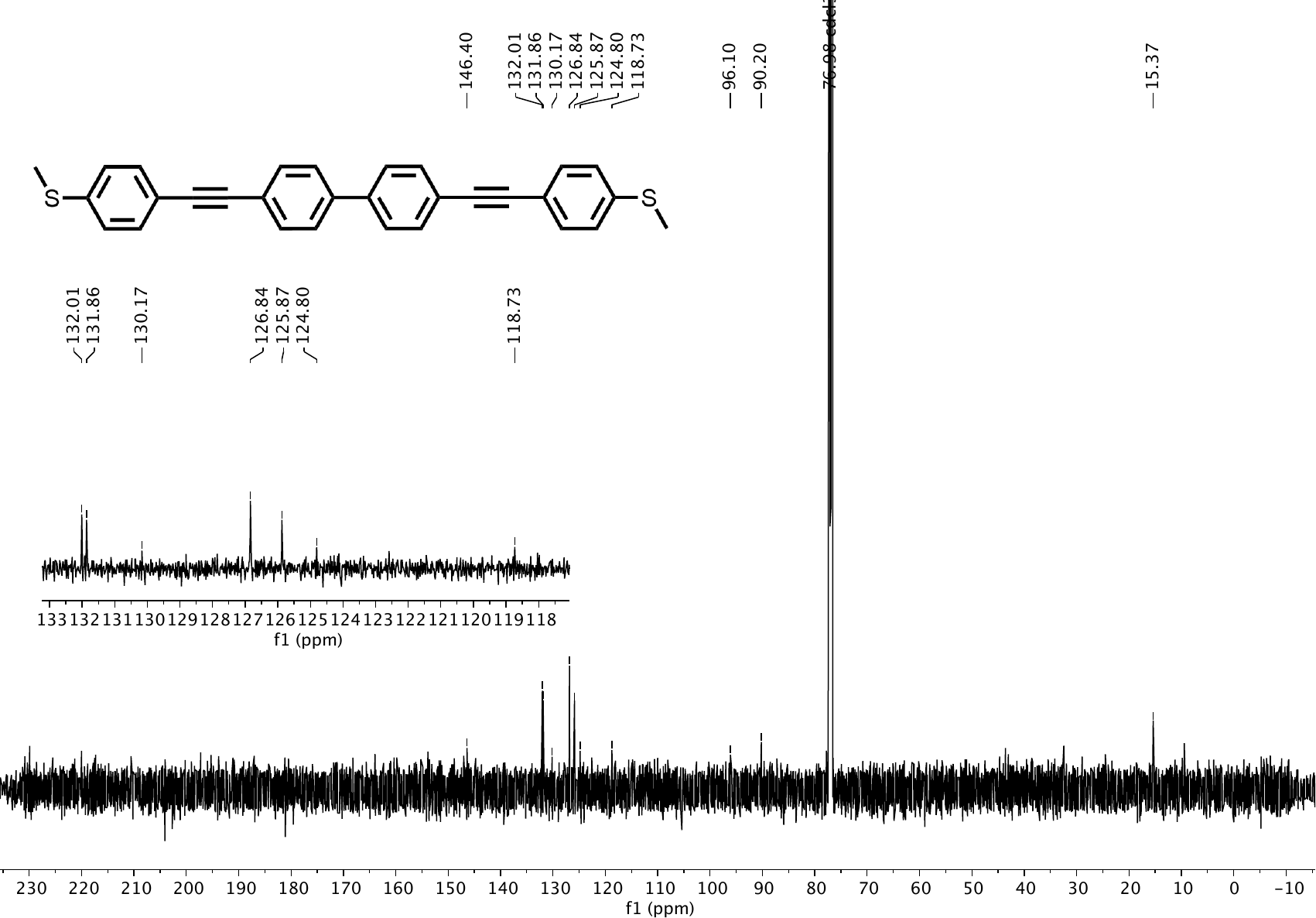}
  \caption{\textsuperscript{13}C NMR (600 MHz) spectra of \textbf{2-SMe} in CDCl\textsubscript{3}.}
  \label{fig:nmr4}
\end{figure}

\begin{figure}[htbp]
  \centering
  \includegraphics[width=0.8\linewidth]{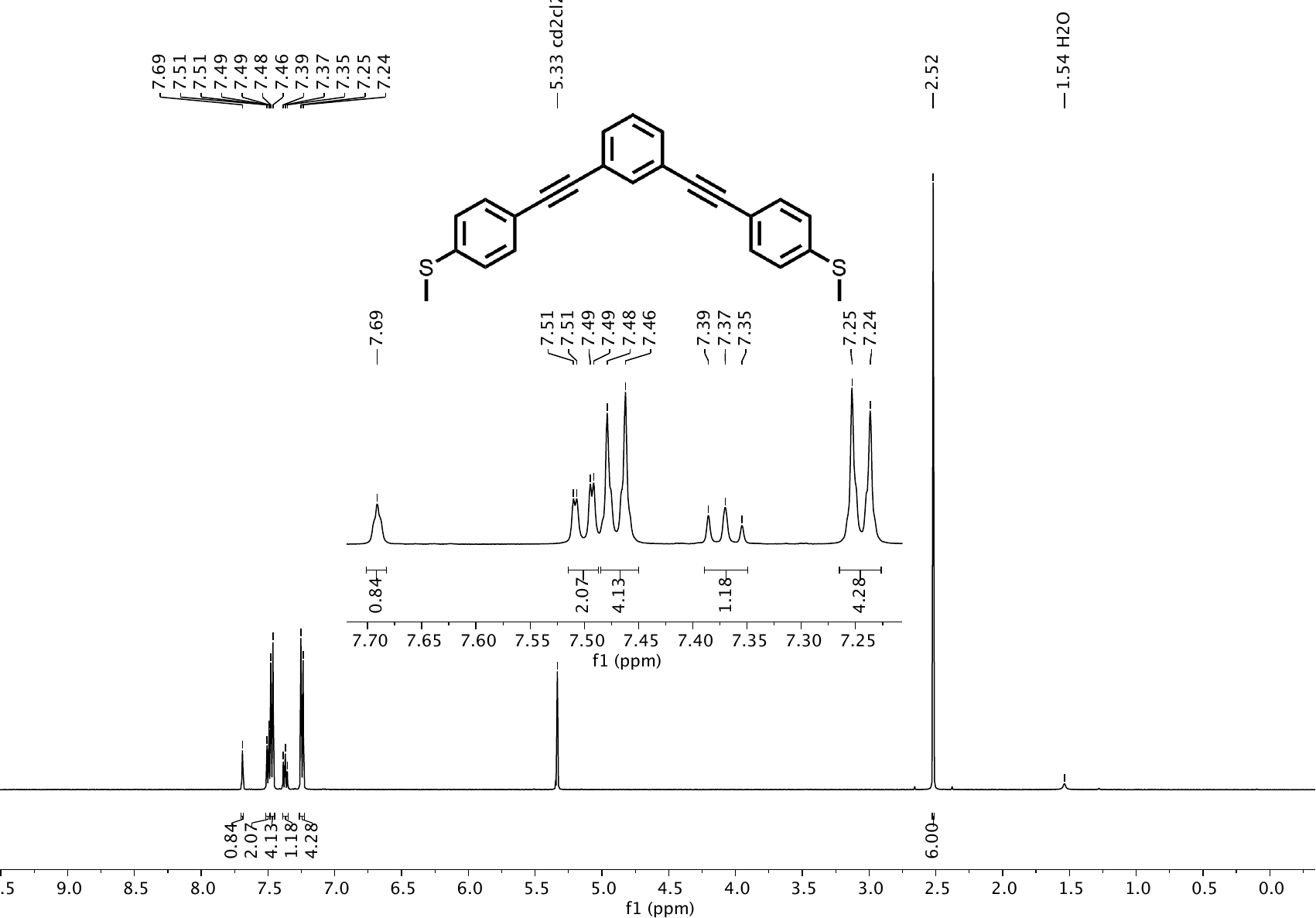}
  \caption{\textsuperscript{1}H NMR (500 MHz) spectra of \textbf{3-meta} in CD\textsubscript{2}Cl\textsubscript{2}.}
  \label{fig:nmr5}
\end{figure}

\begin{figure}[htbp]
  \centering
  \includegraphics[width=0.8\linewidth]{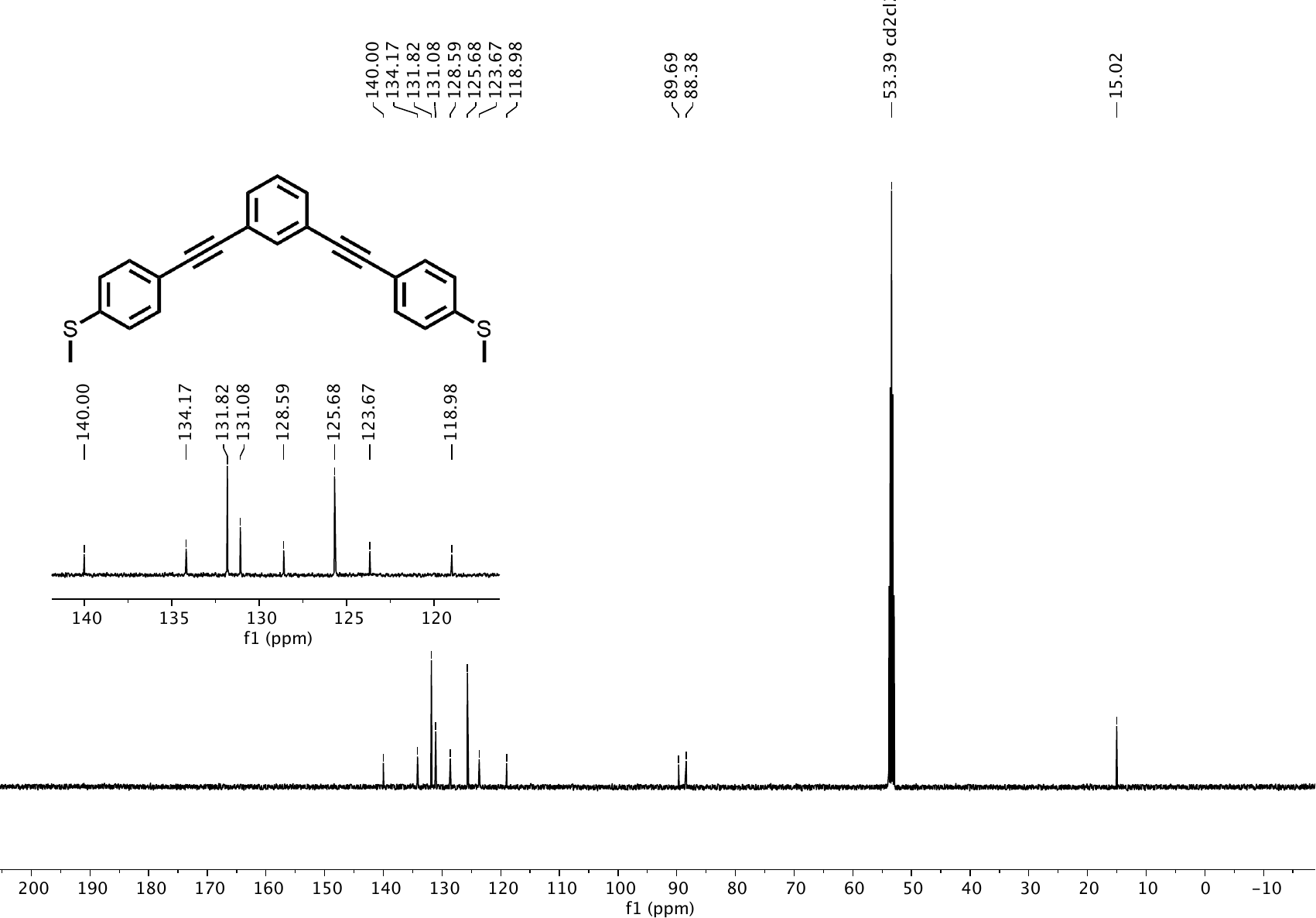}
  \caption{\textsuperscript{13}C NMR (500 MHz) spectra of \textbf{3-meta} in CD\textsubscript{2}Cl\textsubscript{2}.}
  \label{fig:nmr6}
\end{figure}

\newpage
%%%%%%%%%%%%%%%%%%%%%%%%%%%%%%%%%%%%%%%%%%%%%%%%%%%%%%%%%%%%%%%%%%%%%
%% If you are using classical BibTeX rather than biblatex,
%% remove the \printbibliography and uncomment the \bibliograpy one
%%%%%%%%%%%%%%%%%%%%%%%%%%%%%%%%%%%%%%%%%%%%%%%%%%%%%%%%%%%%%%%%%%%%%
\printbibliography
%\bibliography{acs-template.bib}